\documentclass{aa}  

\usepackage{graphicx}
\usepackage{txfonts}
\usepackage{xcolor}

\newcommand{\code}[1]{\texttt{\detokenize{#1}}}

\begin{document}

   \title{The Enhanced Resolution Imager and Spectrograph for the VLT}


\author{R. Davies\inst{1}
\and O.~Absil\inst{11}\fnmsep\thanks{F.R.S.-FNRS Senior Research Associate}
\and G.~Agapito\inst{2}
\and A.~Agudo~Berbel\inst{1}
\and A.~Baruffolo\inst{4}
\and V.~Biliotti\inst{2}
\and M.~Bonaglia\inst{2}
\and M.~Bonse\inst{6}
\and R.~Briguglio\inst{2}
\and P.~Campana\inst{10}
\and Y.~Cao\inst{1}
\and L.~Carbonaro\inst{2}
\and A.~Cortes\inst{8}
\and G.~Cresci\inst{2}
\and Y.~Dallilar\inst{9}
\and F.~Dannert\inst{6}
\and R.~J.~De~Rosa\inst{10}
\and M.~Deysenroth\inst{1}
\and I.~Di~Antonio\inst{3}
\and A.~Di~Cianno\inst{3}
\and G.~Di~Rico\inst{3}
\and D.~Doelman\inst{7}
\and M.~Dolci\inst{3}
\and R.~Dorn\inst{8}
\and F.~Eisenhauer\inst{1}
\and S.~Esposito\inst{2}
\and D.~Fantinel\inst{4}
\and D.~Ferruzzi\inst{2}
\and H.~Feuchtgruber\inst{1}
\and G.~Finger\inst{1}
\and N.~M.~F\"orster~Schreiber\inst{1}
\and X.~Gao\inst{5}
\and H.~Gemperlein\inst{1}
\and R.~Genzel\inst{1}
\and S.~Gillessen\inst{1}
\and C.~Ginski\inst{13}
\and A.~M.~Glauser\inst{6}
\and A.~Glindemann\inst{8}
\and P.~Grani\inst{2}
\and M.~Hartl\inst{1}
\and J.~Hayoz\inst{6}
\and M.~Heida\inst{8}
\and D.~Henry\inst{5}
\and R.~Hofmann\inst{1}
\and H.~Huber\inst{1}
\and M.~Kasper\inst{8}
\and C.~Keller\inst{7}
\and M.~Kenworthy\inst{7}
\and K.~Kravchenko\inst{1}
\and H.~Kuntschner\inst{8}
\and S.~Lacour\inst{12}
\and J.~Lightfoot\inst{5}
\and D.~Lunney\inst{5}
\and D.~Lutz\inst{1}
\and M.~Macintosh\inst{5}
\and F.~Mannucci\inst{2}
\and M.~Marsset\inst{10}
\and A.~Modigliani\inst{8}
\and M.~Neeser\inst{8}
\and G.~Orban~de~Xivry\inst{11}
\and T.~Ott\inst{1}
\and L.~Pallanca\inst{10}
\and P.~Patapis\inst{6}
\and D.~Pearson\inst{5}
\and E.~Pe\~na\inst{10}
\and I.~Percheron\inst{8}
\and A.~Puglisi\inst{2}
\and S.~P.~Quanz\inst{6}
\and S.~Rabien\inst{1}
\and C.~Rau\inst{1}
\and A.~Riccardi\inst{2}
\and B.~Salasnich\inst{4}
\and H.-M.~Schmid\inst{6}
\and J.~Schubert\inst{1}
\and B.~Serra\inst{8}
\and T.~Shimizu\inst{1}
\and F.~Snik\inst{7}
\and E.~Sturm\inst{1}
\and L.~Tacconi\inst{1}
\and W.~Taylor\inst{5}
\and A.~Valentini\inst{3}
\and C.~Waring\inst{5}
\and E.~Wiezorrek\inst{1}
\and M.~Xompero\inst{2}
}

\institute{Max-Planck-Institut f\"ur extraterrestrische Physik, Postfach 1312, 85741, Garching, Germany
\and INAF -- Osservatorio Astrofisico di Arcetri, Largo E. Fermi 5., 50125, Firenze, Italy
\and INAF -- Osservatorio Astronomico d'Abruzzo, Via Mentore Maggini, 64100, Teramo, Italy
\and INAF -- Osservatorio Astronomico di Padova, Vicolo dell'Osservatorio 5, 35122, Padova, Italy
\and STFC UK ATC, Royal Observatory Edinburgh, Blackford Hill. Edinburgh, EH9 3HJ, UK
\and ETH Zurich, Institute of Particle Physics and Astrophysics, Wolfgang-Pauli-Strasse 27, 8093 Zurich, Switzerland
\and Leiden Observatory, University of Leiden, P.O. Box 9513, 2300 RA Leiden, The Netherlands
\and European Southern Observatory, Karl-Schwarzschildstr. 2, 85748, Garching, Germany
\and I. Physikalisches Institut, Universit\"at zu K\"oln, Z\"ulpicher Str. 77, 50937, K\"oln, Germany
\and European Southern Observatory, Alonso de C\'ordova 3107, Vitacura, Santiago, Chile
\and Space Sciences, Technologies, and Astrophysics Research Institute, Universit\'e de Li\`ege, 4000 Sart Tilman, Belgium
\and LESIA, Observatoire de Paris, PSL, CNRS, Sorbonne Universit\'e, Universit\'e de Paris, 5 Place Janssen, 92195 Meudon, France
\and Centre for Astronomy, Dept. of Physics, National University of Ireland Galway, University Road, Galway H91 TK33, Ireland
}


 
\abstract{ERIS, the Enhanced Resolution Imager and Spectrograph, is an instrument that both extends and enhances the fundamental diffraction limited imaging and spectroscopy capability for the VLT.
It replaces two instruments that were being maintained beyond their operational lifetimes, combines their functionality on a single focus, provides a new wavefront sensing module for natural and laser guide stars that makes use of the Adaptive Optics Facility, and considerably improves on their performance.
The observational modes ERIS provides are integral field spectroscopy at 1--2.5~$\mu$m, imaging at 1--5~$\mu$m with several options for high contrast imaging, and longslit spectroscopy at 3--4~$\mu$m, 
The instrument is installed at the Cassegrain focus of UT4 at the VLT and, following its commissioning during 2022, has been made available to the community.}

   \keywords{Instrumentation: adaptive optics -- 
             Instrumentation: high angular resolution -- 
             Instrumentation: spectrographs -- 
             Instrumentation: miscellaneous}

   \maketitle
%

\section{Introduction}
\label{sec:intro}

For one and a half decades, the fundamental near-infrared adaptive optics capability for the VLT was provided by SINFONI and NACO.
SINFONI was installed in 2004 as the combination of the SPIFFI integral field spectrometer \citep{eis03}, which had been previously operated in 2003 as a seeing-limited guest instrument, and the MACAO adaptive optics system \citep{bon04}.
NACO was installed in 2001 as the combination of the CONICA imager and spectrograph \citep{len03} and the NAOS adaptive optics system \citep{rou03}.
With these two instruments working beyond their 10-year design lifetimes, a new facility was required to maintain the diffraction limit capabilities of the telescope into the 2030s, as well as achieve a major enhancement of the performance previously attainable in order to stay competitive.
This need is fulfilled by ERIS, the Enhanced Resolution Imaging Spectrograph.
It is a Cassegrain instrument on UT4 that combines a new imaging camera (NIX), with a refurbishment and upgrade of the integral field spectrometer (SPIFFIER), and a new adaptive optics module that makes use of the Adaptive Optics Facility (AOF; \citealt{ars17,obe18}).

Following the Phase A study, which was concluded in 2012, the project was re-organised and the consortium re-structured to comprise: 
Max-Planck-Institut f\"ur extraterrestrische Physik in Germany (the PI institute, responsible for system engineering, central structure, and integral field spectrometer); 
Istituto Nazionale di Astrofisica in Italy, including the Osservatorio Astrofisico di Arcetri (adaptive optics and warm optics), Osservatorio Astronomico d'Abruzzo (calibration unit), and Osservatorio Astronomico di Padova (control software); 
UK Astronomy Technology Centre in the UK (imager); 
Eidgen\"ossische Technische Hochschule Z\"urich in Switzerland (NIX wheels with their filters and masks);
Sterrewacht Leiden in the Netherlands (phase masks); 
and the European Southern Observatory (detectors and handling tool).

Phase B began at the end of 2014 with a mandate from ESO's Scientific Technical Committee that, given the development landscape, the project should be aiming for a first light not later than 2020.
The Preliminary and Final Design Phases were completed relatively quickly by mid 2017, and SINFONI was decommissioned mid 2019 so that it could be refurbished.
It was, however, only in 2021 that ERIS completed its Preliminary Acceptance in Europe, due to a combination of technical challenges and the impact of the pandemic which led to severe restriction of lab access and curtailed travel.
After shipping, the Assembly, Integration, and Verification of ERIS at Paranal Observatory began within a few weeks of the international borders being re-opened.
Commissioning was carried out during six observing runs in 2022, and progressed well enough that the instrument could be included in ESO's Call for Proposals for Period 111 (for observations beginning in April 2023), which came out in late summer of that year.
The science verification run was performed in December 2022.

\begin{figure*}
\centering
\includegraphics[width=15cm]{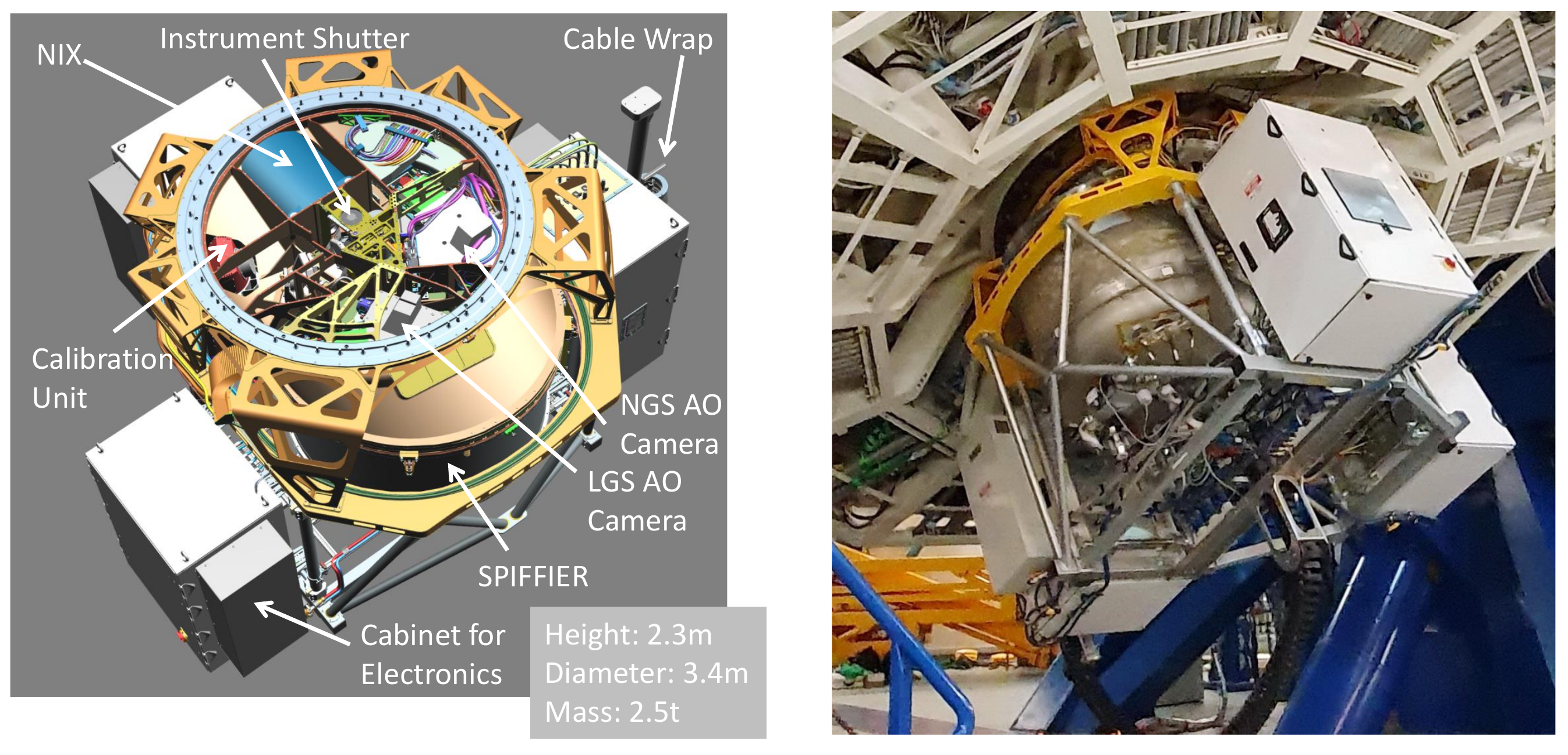}
\caption{Overview of the ERIS instrument and its global characteristics. Left: the CAD drawing with major sub-systems labelled: the SPIFFIER integral field spectrograph, the NIX camera, the Calibration unit, the Central Structure with the wavefront sensor cameras, and the electronics mounted in three cabinets. Right: a view of ERIS mounted at the Cassegrain focus of UT4 at the VLT observatory.}
\label{fig:overview}
\end{figure*}

A description of the ERIS instrument and its major sub-systems follows in Sec.~\ref{sec:description}.
After this, the main science drivers that shaped the design of the instrument are outlined in Sec.~\ref{sec:scidriver}. 
Some early results obtained at different stages of the instrument commissioning are shown in Sec~\ref{sec:commresults} in order to illustrate the capabilities of ERIS in the context of the science drivers, and to compare it's performance to similar results from SINFONI and NACO.

\section{Instrument Description}
\label{sec:description}

ERIS comprises several sub-systems as indicated in Fig.~\ref{fig:overview} and summarised below:
\begin{itemize}
\item
The Adaptive Optics (AO) module, which provides wavefront sensing and works together with the AOF. In addition to high-order correction using a natural guide star (NGS) or a laser guide star (LGS) with a tip-tilt star, it also enables a seeing-enhancer mode and seeing-limited observations.
\item
The Calibration Unit, which allows internal calibration and registration of the wavefront sensors at optical wavelengths, and the two science instruments at 1--2.5~$\mu$m. Longer wavelength calibrations for NIX are performed on sky.
\item
The integral field spectrometer SPIFFIER, which operates at 1--2.5~$\mu$m, provides $64\times32$ spatial pixels (at $12.5\times25$~mas over $0\farcs8\times0\farcs8$; $50\times100$~mas over $3\farcs2\times3\farcs2$; and $125\times250$~mas over $8\farcs0\times8\farcs0$), and two spectral resolutions ($R\sim5000$ over full bands, or $R\sim10000$ over half-bands).
, 
\item
The imager NIX, which operates at 1--5~$\mu$m, provides two pixel scales (13~mas covering $26\farcs4\times26\farcs4$, and 27~mas covering $55\farcs4\times55\farcs4$) for imaging with broad and narrow band filters, as well as focal and pupil plane masks for high contrast imaging, and a long slit for L-band spectroscopy at $R\sim900$.
\end{itemize}

These sub-systems are integrated in and around a central structure which also provides the interface to the Cassegrain adaptor/rotator of the telescope.
Surrounding this structure is a framework holding three electronics cabinets which have adjustable mounts that facilitate balancing the instrument's centre of gravity such that the moment vector parallel to the optical axis becomes smaller than the required 500~nm at any telescope orientation. A fourth cabinet, containing the real-time computer for the AO system (Sec.~\ref{sec:AO}) and the local control units is on the floor of the dome.

\begin{figure}
\centering
\includegraphics[width=\hsize]{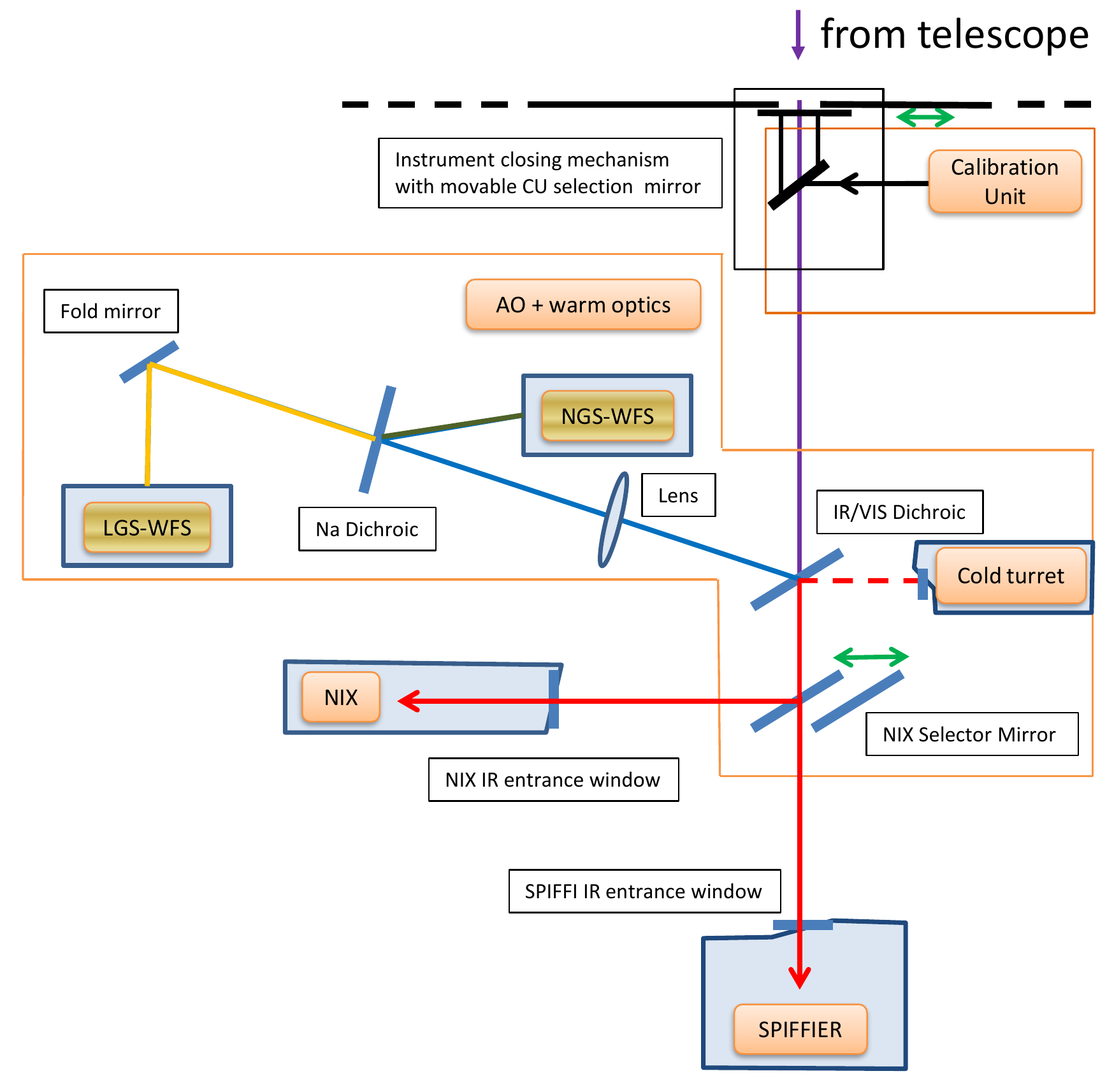}
\caption{Concept for the optics in the central structure. The light from the telescope can be blocked by the instrument closing mechanism, which instead folds in light from the Calibration Unit. The optical path then goes directly to SPIFFIER, passing only the IR/VIS dichroic that splits light off to the wavefront sensors. The path can be redirected to NIX with a movable fold mirror (the background from which is kept low via a turret extension of the SPIFFIER cryostat). The AO optical path is split by another dichroic which transmits light from the LGS, and reflects the rest to the NGS wavefront sensor.}
\label{fig:censtruct}
\end{figure}

\begin{figure}
\centering
\includegraphics[width=\hsize]{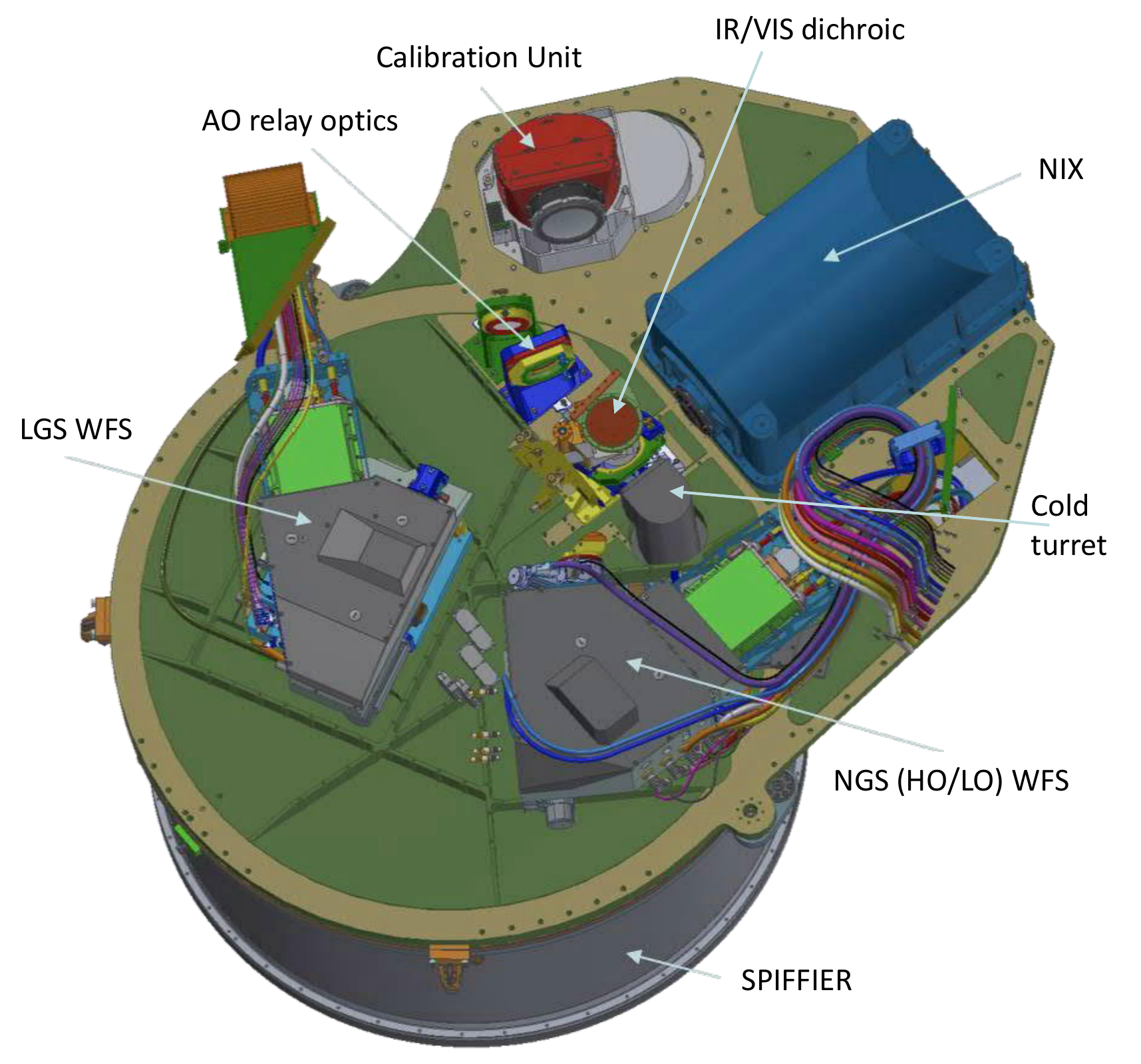}
\caption{The optical bench of the `Warm Optics', which fits into a volume only 30~cm high. The IR/VIS dichroic reflects the visible light to the AO relay optics. The next element is a notch filter that transmits the laser light, which is then reflected by a fold mirror to the LGS WFS. The rest of the visible light is reflected to the NGS WFS.}
\label{fig:AO_WO}
\end{figure}

Within the central structure are a small number of additional opto-mechanical elements and mechanisms to direct the light accordingly.
The concept for these `Warm Optics' is illustrated in Fig.~\ref{fig:censtruct}, and their layout is shown in Fig.~\ref{fig:AO_WO}. 
The first is the instrument closing mechanism. 
When moved in, the external light path into the instrument is blocked, and instead light from the calibration unit is directed towards the other sub-systems.
Below this unit the height inside the central structure is limited to 30~cm.
This constraint arises in order to avoid the need for re-imaging optics in front of the instruments, by increasing the back focal distance of the telescope from 25~cm to 50~cm.
The most important impact is on the CaF$_2$ IR/VIS dichroic which picks off the optical wavelengths at 0.4--1.0~$\mu$m and directs them to the wavefront sensors (WFSs).
In order to position the WFSs within the available volume, this dichroic is tilted by 45$\degr$. Its rear surface has a small curvature and wedge to correct the astigmatism introduced on the transmitted beam; but there remains a small anamorphic magnification which is seen as a distortion in the NIX field as described in Sec.~\ref{sec:niximaging}.
Following the reflected beam, an additional notch filter allows transmission of the laser light at the 589~nm wavelength of the Na\,I\,D line to the LGS WFS while other optical wavelengths are reflected to the NGS WFS. 
The last component of the Warm Optics is a selector mirror. When moved out, light from the telescope enters directly into SPIFFIER, only passing the IR/VIS dichroic en route. When moved in, the light is redirected to NIX.
In order to avoid additional thermal background from the central structure being reflected into NIX by the back side of the IR/VIS dichroic, a cold turret periscope extension to the SPIFFIER cryostat has been added so that NIX sees instead a cold surface here.

\subsection{Adaptive Optics}
\label{sec:AO}

\begin{figure}
\centering
\includegraphics[width=\hsize]{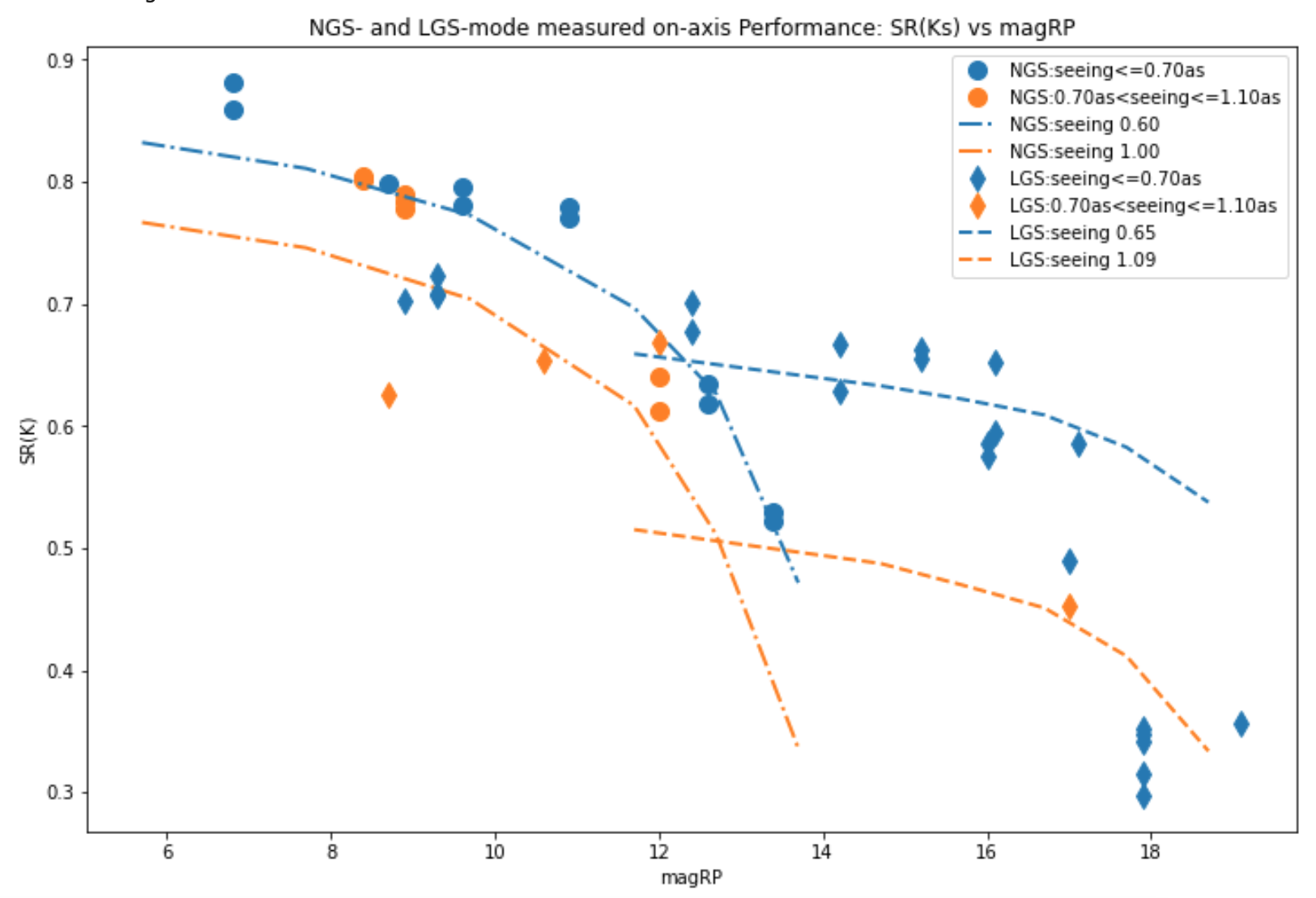}
\caption{Simulated and measured on-axis K-band performance of the AO system for NGS and LGS modes, as a function of guide star magnitude in a dark sky. At the very faint end, excess background leads to a lower performance than predicted; and significant lunar illumination will also have an impact. Performance in other wavebands can be inferred from the associated wavefront error. For estimations of specific situations and conditions, the reader is referred to the online Exposure Time Calculator. More information about the AO system and a detailed breakdown of its performance are given in Riccardi et al. (in prep).}
\label{fig:aoperf}
\end{figure}

The AO module is described in detail by Riccardi et al. (in prep), with an outline of the design given in \cite{ric18} and some initial results reported in \citep{ric22}. Only a brief outline is given here from the observer's perspective, with a simple overview of the on-axis performance in Fig.~\ref{fig:aoperf}.
In addition to the wavefront sensors, the real-time computations make use of ESO's standard platform for adaptive optics real time applications, SPARTA \citep{sua12}. The wavefront correction is applied to the 1170 actuator Deformable Secondary Mirror (DSM) that is part of the AOF. Any one of the lasers in the 4LGSF \citep{hac11} can be used to generate a LGS when required.

The bandpasses of the acquisition camera and NGS WFS are well matched to the Gaia blue ($G_{BP}$) and red ($G_{RP}$) bandpasses \citep{gai16} respectively. Hence the Gaia Data Release 3 catalogue \citep{gai22} -- for which accurate magnitudes, positions and proper motions are known for a large number of stars -- has been selected as the default catalogue from which to specify AO guide stars for ERIS.

When working with a natural guide star, the AO system uses a Shack-Hartmann WFS sensitive in the range 600-1000~nm.
It is mounted on a linear stage, which is used together with a rotating periscope, so that the guide star can be picked up anywhere within a $59\arcsec$ radius patrol field. This allows off-axis guiding, although the atmospheric anisoplanatism will reduce the performance at the science target if it is offset from the guide star.
An acquisition camera, with a $15\arcsec$ field of view and working at 400-580~nm, is mounted in the same unit, and used to centre the guide star in the 2.0$\arcsec$ field of view of the WFS. The limiting magnitude for this is set to $G_{BP} = 19$~mag in order to avoid overly long integration times.
A $40\times40$ lenslet array in the NGS WFS is used to enable high-order correction at frequencies of 300--1000~Hz.
In median seeing, it can provide a Strehl ratio close to 80\% in K-band for guide stars with magnitudes in the range $1 \le G_{RP}~(mag) \le 10$ with performance tailing off to $\sim50$\% at the set limit of $G_{RP}\sim13$~mag.

Sky coverage is increased significantly when using the LGS WFS. But use of the LGS is specified to zenith distances of only 60$\degr$. As such, coverage of the 5.4$\degr$ region around the south celestial pole is only possible with NGS-AO, which can be used to the 70$\degr$ pointing limit of the VLT.
The LGS WFS has a similar design, also with a $40\times40$ lenslet array, but with a $4.4\arcsec$ field of view. It is limited to on-axis use and has instead a linear stage to allow for focussing, in order to track the line-of-sight distance to the mesospheric sodium layer over the range 80--200~km.
The LGS WFS is used in conjunction with the NGS WFS which, for this mode, is reconfigured with a $4\times4$ lenslet array to provide tip-tilt as well as enable truth sensing for focus and other low-order terms. The star can be picked up anywhere within the $59\arcsec$ radius patrol field, although again the performance for stars that are far off-axis will decrease due to anisokinetism. 
In median seeing, and with a nearby tip-tilt star in the magnitude range $7 \le G_{RP}~(mag) \le 16$, LGS AO can achieve Strehl ratio around 60\% in the K-band. At fainter magnitudes the performance falls to about 30\% at the set limit of $G_{RP} = 18$~mag.
The performance at the faint end is dependent on the lunar illumination, which can add a significant background. In dark conditions and with good seeing, the loop has been closed and stable using a star one magnitude fainter than the formal limit.

The AO system provides two more levels of correction. 
Using the LGS WFS without the NGS WFS corresponds to the seeing-enhancer mode \citep{dav08}. In this case, jitter compensation is provided by the VLT's guiding capability which can run at a frequency of up to 30~Hz. Without the truth sensor, the focus term from the telescope's active optics is instead used as the feedback for the LGS focus stage tracking, to compensate for the changing distance to the sodium layer. With this mode of operation, one can typically achieve a resolution of 0.2--0.3$\arcsec$.
Finally, seeing-limited operation does not use either WFS. In this mode, the AO system simply applies a nominal (flat) shape to the DSM.

In addition to providing high-order correction while field tracking, the AO system can also operate when differential tracking between the NGS WFS and the telescope is required. This can occur for two reasons.
The first is pupil tracking, which is essential for the high contrast modes of NIX (described in Sec.~\ref{sec:nix}), where angular differential imaging \citep[ADI; ][]{Marois_2006} is the main technique used to process and combine the individual exposures.
The other case concerns the non-sidereal modes. If the science target and the AO reference source are the same object, then this is achieved trivially by adjusting the tracking of the telescope according to an ephemeris file provided by the observer, and correcting any deviations via the AO system.
If, instead, there is differential motion between the AO reference source and the science target (most typically the AO reference is a star, and the science target is a solar system object), then the relative NGS WFS pointing must be continually updated according to an ephemeris file. This relies on precise calibration of the stages of the NGS WFS, and correction of any deviations can only be done during data processing. Verification of this mode was performed by observations of 216~Kleopatra, a highly elongated asteroid.
During the observations, it was moving at a non-sidereal speed of about 25\arcsec/hr, and the correction was done with LGS-AO using a tip-tilt star 50\arcsec\ away. The image was stabilised to a precision of about 10~mas, consistent with anisokinetism.

\subsection{Calibration Unit}

\begin{figure}
\centering
\includegraphics[width=\hsize]{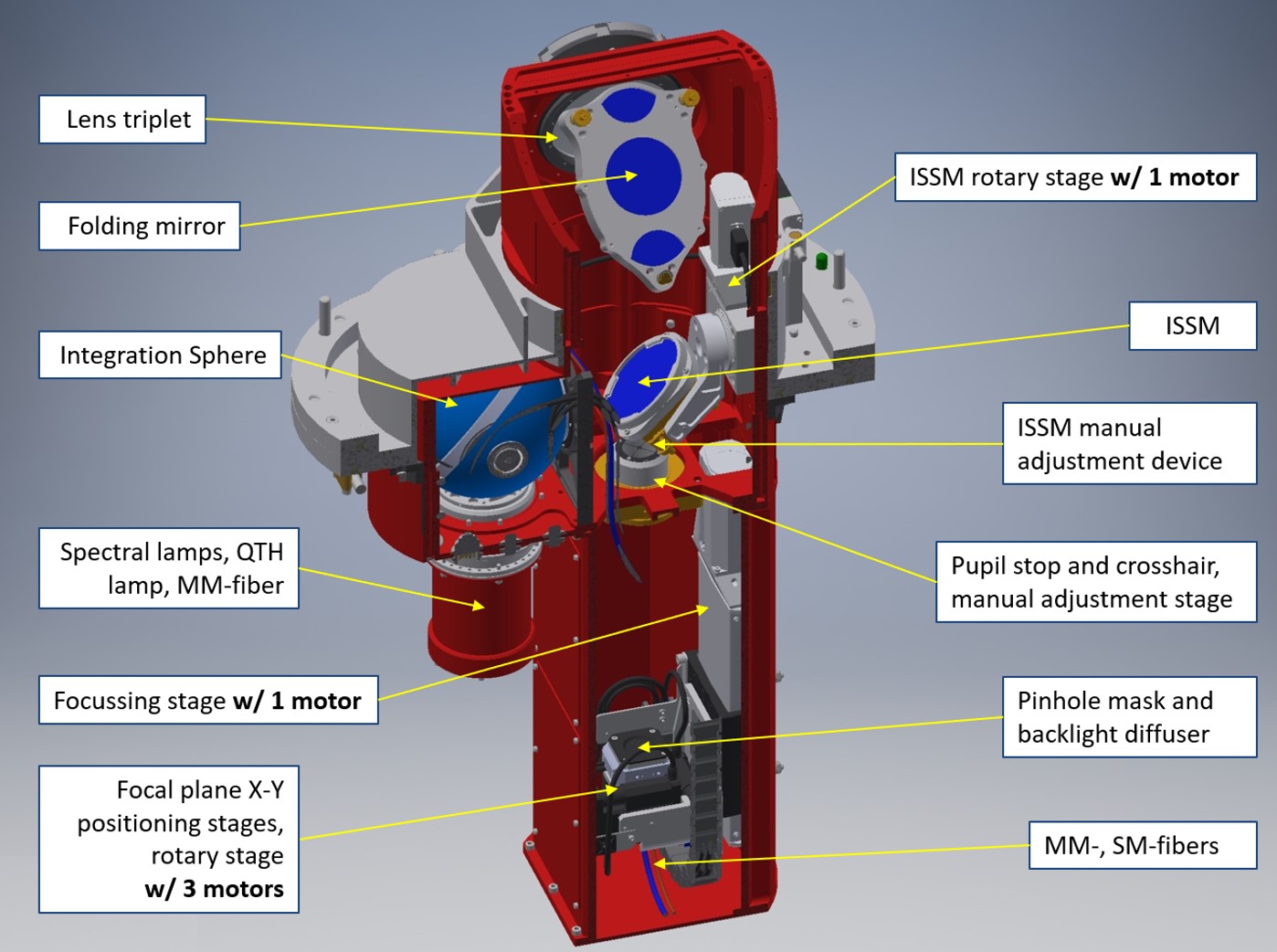}
\caption{Drawing of the calibration unit main bench indicating the primary components, which is used only in the orientation shown (calibrations are taken with the telescope at zenith). The triplet lens interfaces to the central structure. Below this, the ISSM enables selection of the integrating sphere (for flatfield and arclamp calibrations) and the back-illuminated mask (which allows projection of point/extended sources and slitlets across the field).}
\label{fig:cumb}
\end{figure}

The Calibration Unit (CU) enables calibration of the AO system at optical wavelengths, as well as NIX and SPIFFIER at 1--2.5~$\mu$m. In order to avoid compromising its performance at those core wavelength ranges, and because of opto-mechanical constraints, it does not provide 3--5~$\mu$m calibrations for the longer wavelength camera in NIX. Instead, that camera, including the L-band long-slit spectroscopy mode, must be calibrated on-sky.

The CU main bench, shown in Fig.~\ref{fig:cumb}, is opto-mechanically interfaced to the Warm Optics in the central structure. It consists of two main sub-units.
The first is the integrating sphere, which can be directly illuminated by two sets of light sources:
\begin{enumerate}[(i)]
\item A quartz-tungsten halogen lamp for flatfielding the detectors in NIX and SPIFFIER, which can be used with all the 1--2.5~$\mu$m filter and grating configurations.
\item A set of four pencil-ray lamps (Ne, Xe, Kr, and Ar) which enable wavelength calibration at 1--2.5~$\mu$m in SPIFFIER. They are used in different combinations depending on the grating configuration.
\end{enumerate}
These can be accessed by inserting the integrating sphere selector mirror (ISSM), which allows uniform illumination from the integrating sphere to propagate out of the CU and into the Central Structure.
When the ISSM is moved out, the other main sub-unit comes into play. It includes a pupil stop imitating that of the telescope; and has a mask hosting sets of slits and pinholes which are back-illuminated by an Energetiq laser-driven light source that provides an almost flat spectrum over 0.4--2.4~$\mu$m. It is physically located in one of the electronics cabinets, and the light is transmitted to the CU with multi-mode and single-mode fibres. Due to the range of wavelengths that any single fibre can transmit, different fibres are used for the visible and near-infared regimes.
A variable neutral density filter allows adjustment of the light intensity across four orders of magnitude.
The focussing stage of this sub-unit allows one to take account of the range of distances to the sodium layer for LGS calibrations. The mask is on a stage allowing translational offsets as well as rotation, to provide maximum flexibility for aligning the various sources or positioning them anywhere in the focal plane. The mask includes:
\begin{enumerate}[(a)]
\item Two sets of three slits, for calibrating the spectral curvature and alignment of the slitlets in SPIFFIER. These slits are aligned perpendicular to the slitlets and replace the `north-south' calibration of SINFONI.
\item A diffraction limited source for aligning the instruments and wavefront sensors, measuring differential flexure, and deriving non-common path aberrations between them. This is also used for AO calibration and testing of the NGS WFS.
\item Extended ($0.5\arcsec$, $1.0\arcsec$, and $1.5\arcsec$ diameter) sources for AO calibration of the LGS WFS, and also the NGS WFS when correcting on extended objects.
\end{enumerate}

\subsection{SPIFFIER Integral Field Spectropgraph}
\label{sec:spiffier}

\begin{figure}
\centering
\includegraphics[width=8cm]{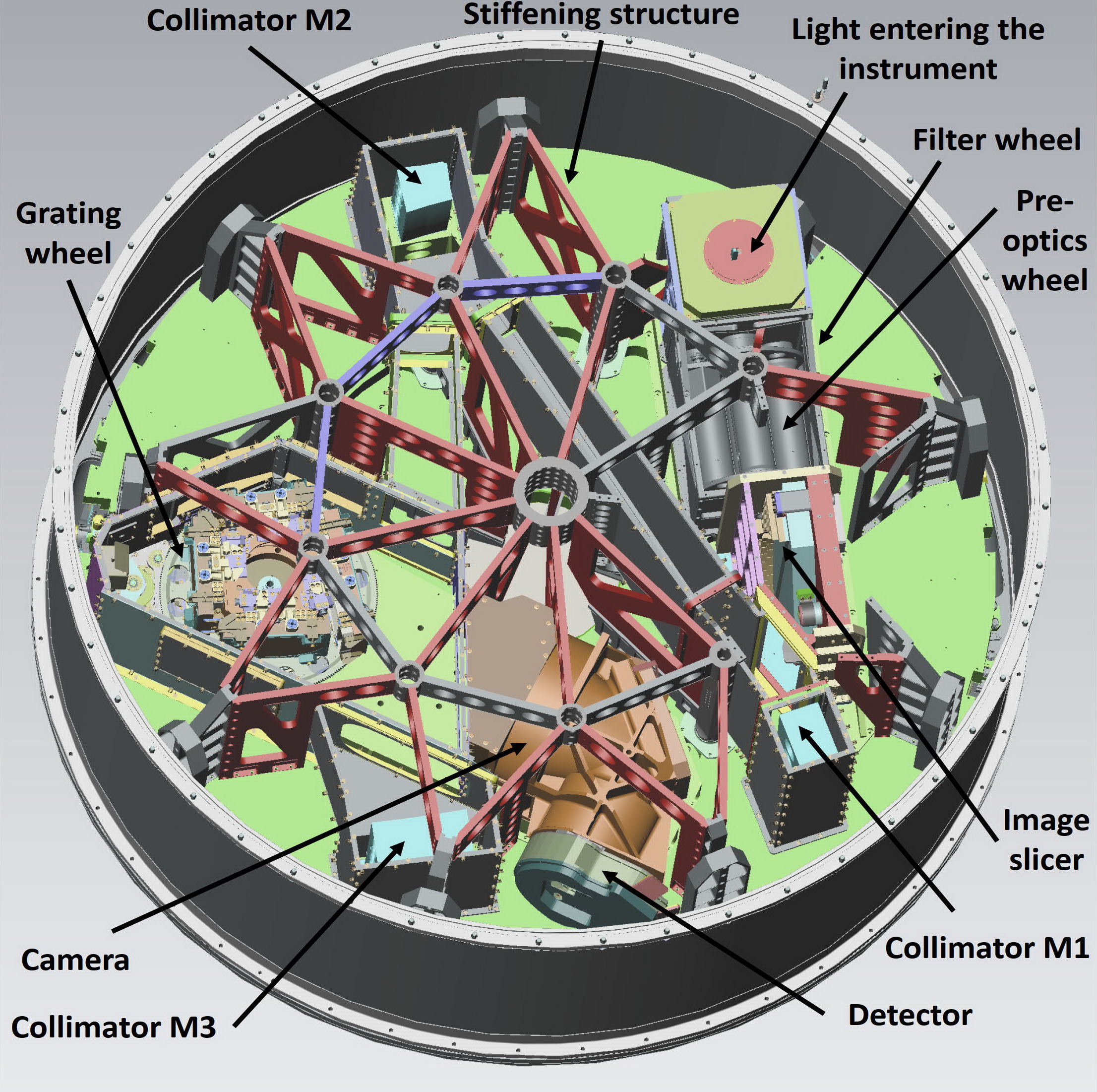}
\caption{Drawing of the inside of the SPIFFIER cryostat, showing the optical path and locations of the main cryogenic opto-mechanical components. The cryostat is cooled by liquid nitrogen. In order to illustrate these better some parts of the housings and stiffening structure have been omitted.}
\label{fig:spiffier}
\end{figure}

SPIFFIER is an upgrade and refurbishment of SPIFFI, the integral field spectrometer in SINFONI, to improve the throughput as well as the image and spectral quality. Opto-mechanically and functionally, though, the instrument remains basically the same as shown by the layout in Fig.~\ref{fig:spiffier}, and is cooled by a liquid nitrogen bath. 
Light enters the instrument from above, the incoming beam is collimated before passing first through the filter wheel to select the waveband and then a $\sim6$~mm cold stop which includes a central obscuration. After this, the pre-optics unit enables selection of the pixel scale and focusses the light on the image slicer, which re-arranges the square field into a long slit format as described in \citet{eis03}.
A three-mirror anastigmat then collimates the beam so that it can be dispersed by one of the gratings on the grating wheel. The light is finally re-imaged in the camera and sensed by the detector.

\begin{figure}
\centering
\includegraphics[width=\hsize]{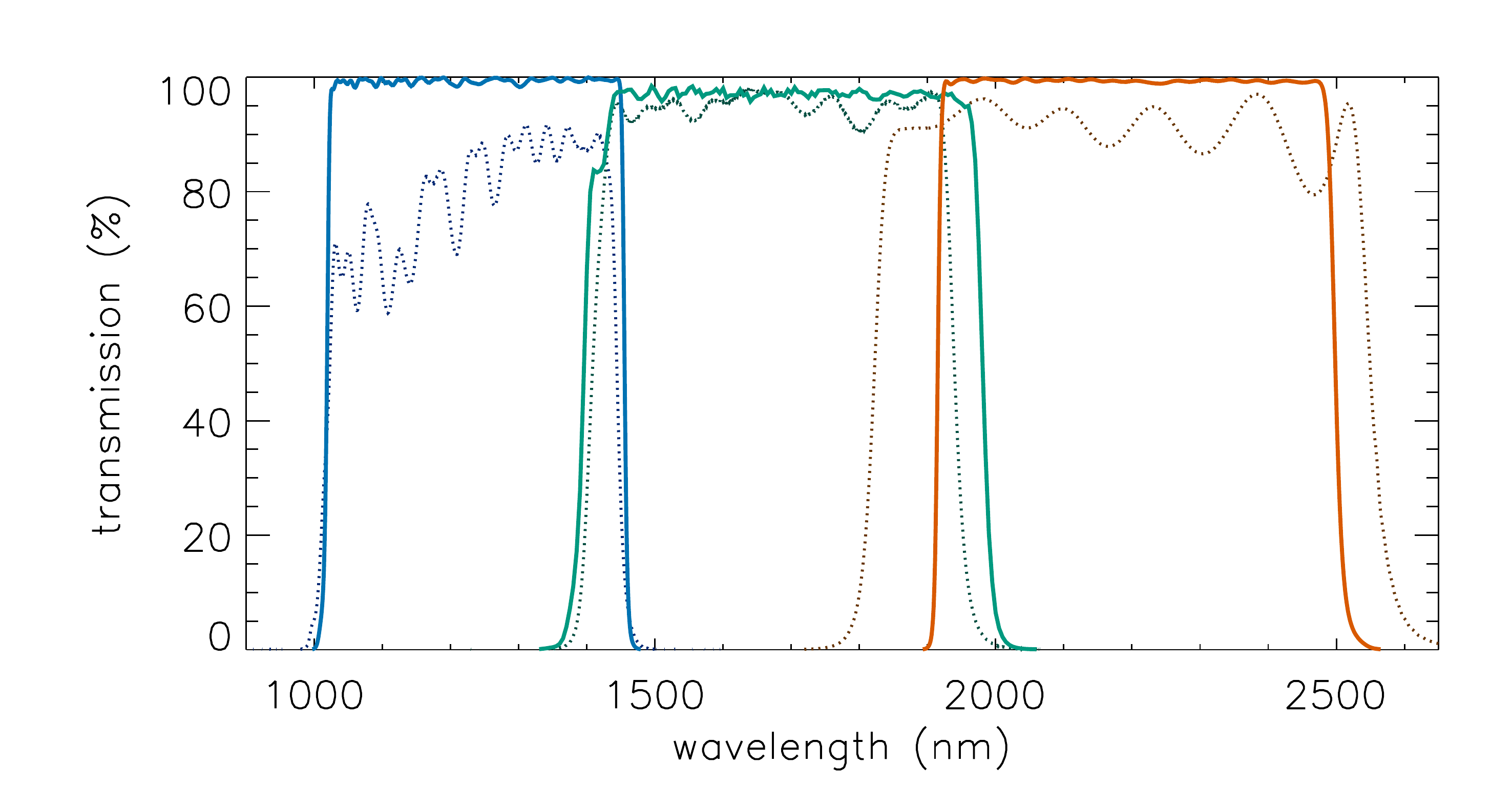}
\caption{Filter throughputs in SPIFFIER for the low resolution gratings. The dark dotted lines denote the filters used before the refurbishment; the thicker solid blue, green, and red lines for the J, H, and K bands are afterwards. This change alone leads to increases in throughput by 5-20\%.}
\label{fig:spiffierfilters}
\end{figure}


\begin{figure*}
\centering
\includegraphics[width=14cm]{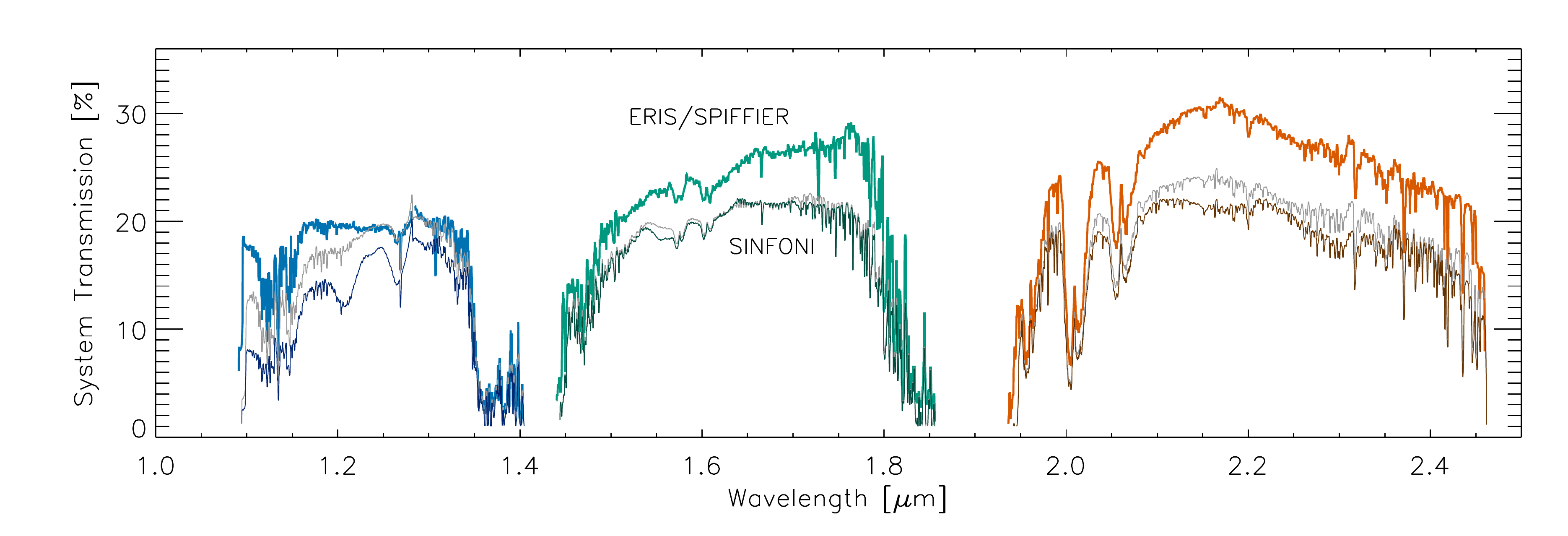}
\caption{Total transmission -- including telescope, atmosphere, and detector quantum efficiency -- for ERIS/SPIFFIER (thick curves) with a comparison to the equivalent curve for SINFONI (thinner darker curves, for pre-2016; grey curves represent the situation after the upgrade in 2016 when replacement of the filters led to an increase of the transmission). The data are averaged from several observations of the spectrophotometric standard star LTT~3218 in seeing-limited conditions, and were extracted in a large aperture. The transmission curves do not reflect gains due to the improved spectral line profile or AO performance.}
\label{fig:SPIFFIERthruput}
\end{figure*}

Operationally, the main difference is that the lower resolution H+K grating has been replaced with a higher resolution $R\sim10000$ grating. This covers half a band at a time, and can be used in any of three fixed settings in each of the J, H, and K atmospheric windows.
The other three gratings, which each cover a full single band at $R\sim5000$, have also been replaced in order to improve the spectral line profile.
The pre-optics have been re-coated, but are otherwise unchanged so the three pixel scales remain the same: $12.5\times25$~mas, $50\times100$~mas, and $125\times250$~mas, where the long side corresponds to the width of the slitlets and the short side the size of the detector pixels.
These provide fields of view of $0\farcs8\times0\farcs8$, $3\farcs2\times3\farcs2$, and $8\farcs0\times8\farcs0$.
The filters have been replaced, to increase the throughput as shown in Fig.~\ref{fig:spiffierfilters}.
And the detector has been replaced with a new $2048\times2048$~pixel Teledyne HAWAII 2RG device which has excellent cosmetic quality, having only one small cluster of bad pixels in slitlet~16.
It is used in a slow read mode only, with characteristics as reported in Table~\ref{tab:detectors}.
Following an exposure at half of the full well depth, the persistence is only to 2-3~e- after 10~mins and $<1$~e- after 30~min. As such, the persistence performance is much better, and more uniform, than the previous detector.
The combination of the changes described above has led to a substantial increase in the total system throughput, as illustrated in Fig.~\ref{fig:SPIFFIERthruput}.

\begin{figure}
\centering
\includegraphics[width=\hsize]{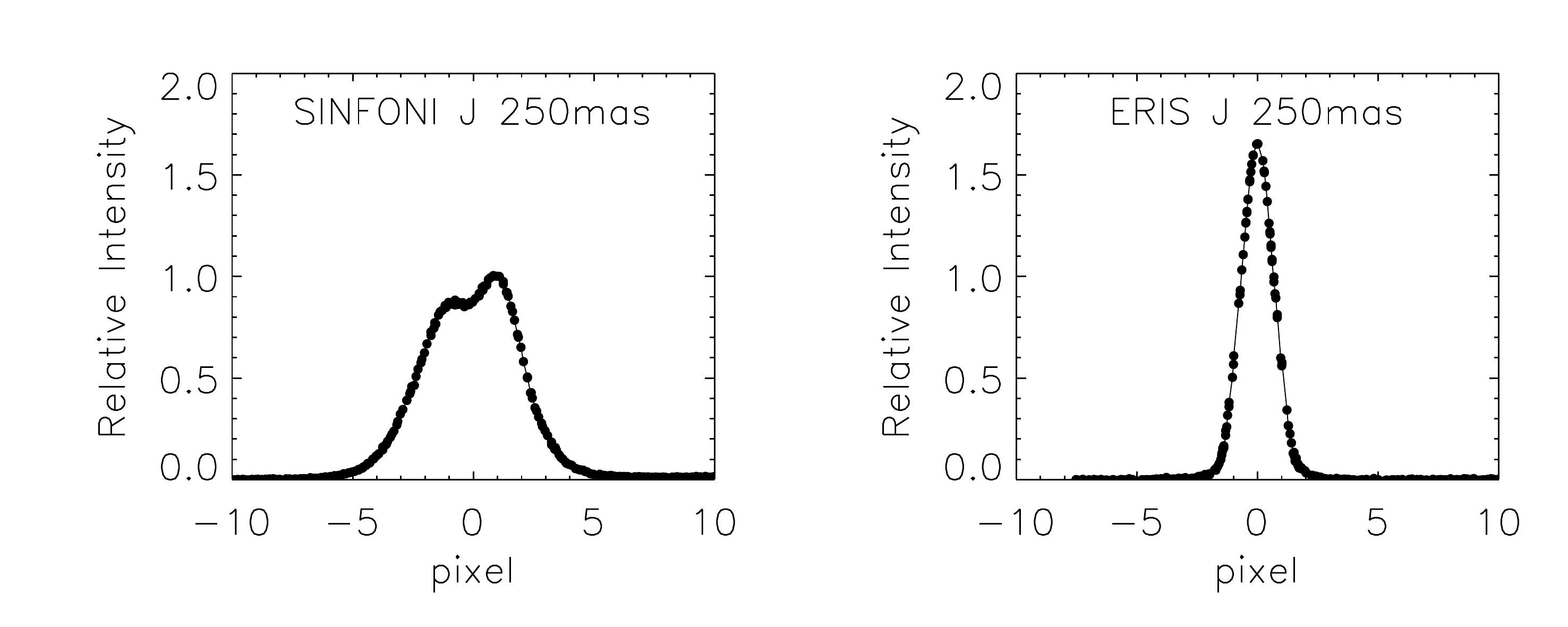}
\caption{Spectral profile of a line near the centre of the J-band with the 250~mas pixel scale. Left: as it was in SINFONI. Right: after replacement of the mirrors and gratings, as it is in ERIS. The profiles have been normalised to the same total flux, showing that not only is the resolution improved, but also the sensitivity to unresolved lines. This effect is more significant at shorter wavelengths and smaller pixel scales.}
\label{fig:Jprofile}
\end{figure}

\begin{figure}
\centering
\includegraphics[width=8cm]{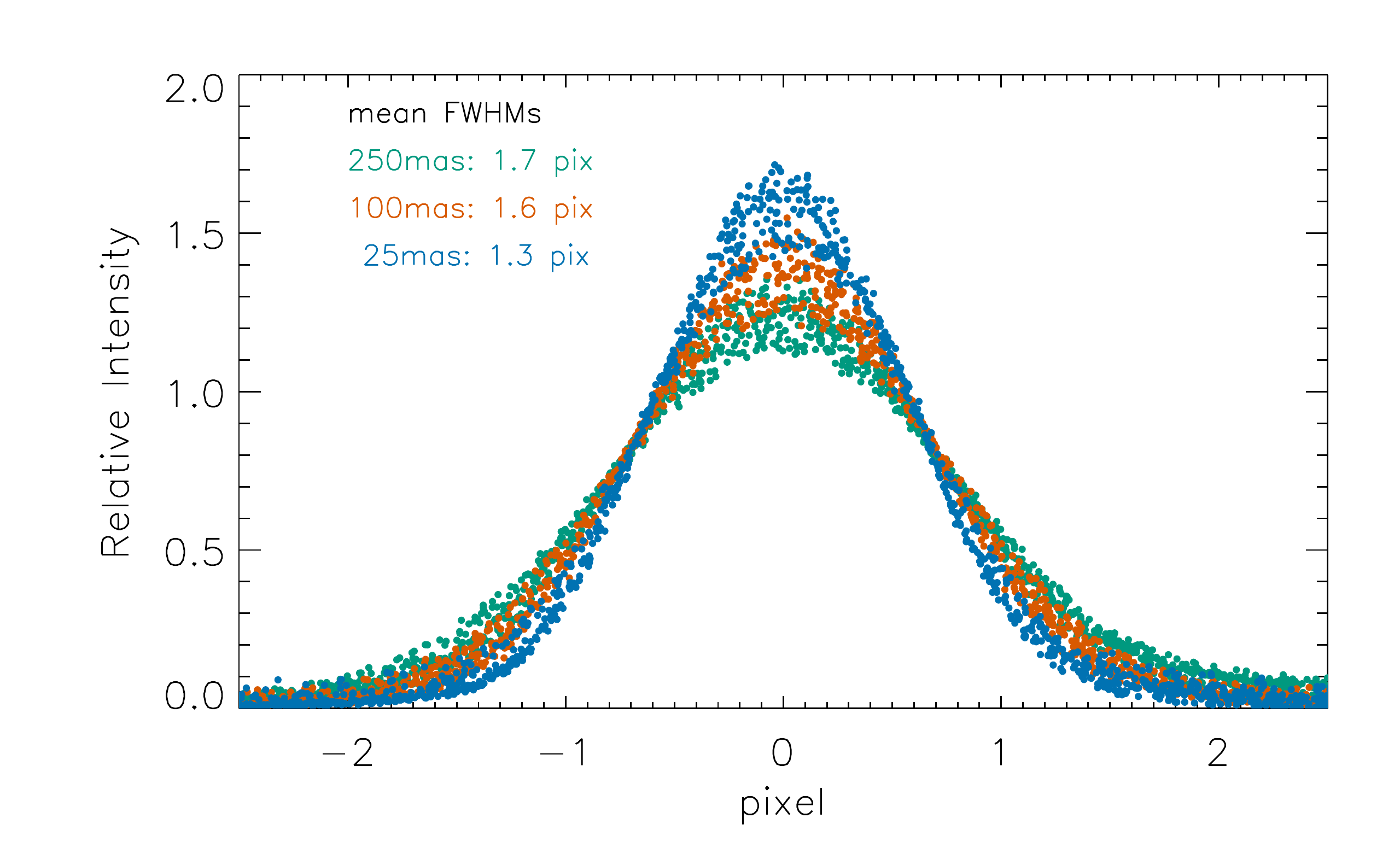}
\caption{Spectral profiles in all the grating configurations of SPIFFIER, normalised by total flux. Each colour corresponds to a pixel scale (250~mas, 100~mas, and 25~mas in green, red, and blue respectively) because this matches the largest variation when measured in pixels. For each pixel scale (or colour) there are 12 measurements corresponding to band (J, H, K) and grating (low, short, middle, long). In all cases, the profiles are smooth and well represented by a Gaussian.}
\label{fig:specprof}
\end{figure}

\begin{figure*}
\centering
\includegraphics[width=14cm]{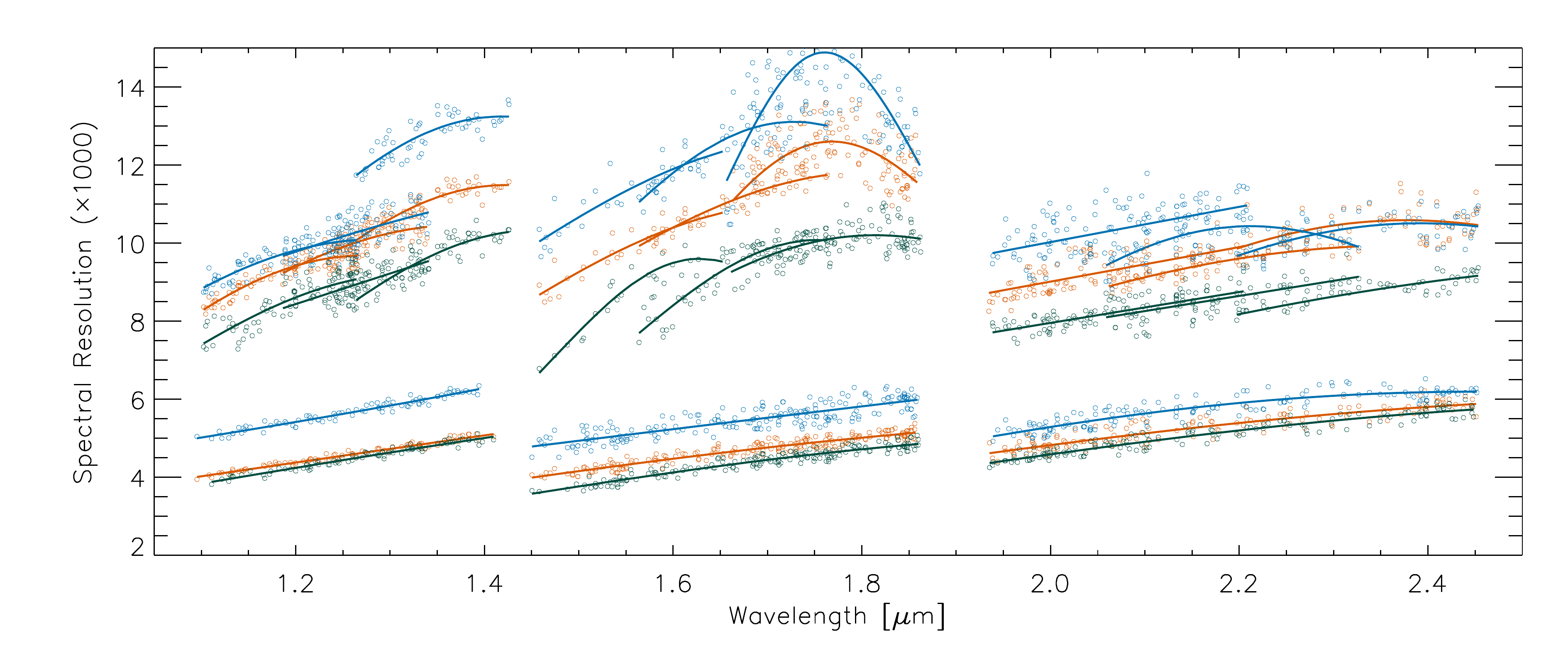}
\caption{Spectral resolution in SPIFFIER for every band, grating, and pixel scale. The data were measured from arclamp exposures, as reported by one of the pipeline products. The pixel scale is indicated by color (250~mas in green, 100~mas in red, 25~mas in blue). The points for each configuration have been fitted as a quadratic function of wavelength.}
\label{fig:specres}
\end{figure*}

Perhaps most importantly the refurbishment of the optics has vastly improved the image and spectral quality. Previously the line spectral profiles were rather complex and varied with wavelength and pixel scale.
In the K-band this led to wings on the line profiles, while in the J-band it meant the lines were double-peaked, as shown in the left panel of Fig.~\ref{fig:Jprofile} (obtained by combining measurements taken with the grating wheel rotated slightly each time in order to take `babysteps' and oversample the profile).
The cause of this was reported as a combination of residual diamond turning marks on the mirrors \citep{geo16} and cryogenic deformation of the gratings due to their lightweighting \citep{geo17}.
Replacing these optics, in particular using new gratings based on Zerodur blanks with no lightweighting, has fulfilled the predictions made in terms of line profile, which are now well represented by a Gaussian with a FWHM in the range 1.3--1.9~pixels. 
As shown in Fig.~\ref{fig:specprof}, there is some variation in line width, with the strongest dependency on pixel scale.
In normal operations the spectral undersampling of the lines means that the measured resolution is slightly lower, and also varies (between that provided by the grating and that due to Nyquist sampling) depending on the exact location of the line centre with respect to the pixel grid.
The spectral resolutions achieved for the various grating configurations are therefore shown in Fig.~\ref{fig:specres}.

The larger mass of the new gratings has led to an increase in spectral flexure as a function of rotator angle and telescope altitude. Measurements indicate that, if uncorrected, this can be up to several pixels. As a result, the pipeline provides a correction to the wavelength calibration (which is only done at zenith, using the arc lamps) based on the atmospheric OH lines. This follows the method proposed by \citet{dav07} and the implementation done for KMOS \citep{dav13}: after a preliminary reconstruction, the offset is measured and applied to the calibration, and the data reconstructed again. This avoids additional interpolations of the science data.

Finally, the motors and instrument control system for ERIS have also been completely replaced. 
This is now based on one Beckhoff Programmable Logic Controller (PLC) for each subsystem \citep{kie14}. 
Each PLC hosts various numbers of analogue inputs (sensor signals) and outputs (control signals), digital inputs (status signals) and outputs (switch signals), as well as motion controllers for the motorised functions \citep{pop14}. 
These mechanisms include the filter wheel, pre-optic wheel, and grating wheel, for which new Phytron motors have been selected to provide similar torques to the previous ones.
A Siemens S7 PLC provides the cryogenic and vacuum control for the cryostats of both NIX and SPIFFIER. 
Each PLC uses the Open Platform Communication Unified Architecture communication protocol with the instrument control software, and software libraries developed by ESO are used to command the devices from the user interface.

\begin{table}[h]
\caption{Basic Detector Characteristics}
\label{tab:detectors}
\centering
\begin{tabular}{llccc}
\hline\hline
 && SPIFFIER & NIX & NIX \\
 && slow & slow & fast \\
 \hline
 min. exp.$^a$  & e- & 1.6 & 1.97 & 0.034 \\
 gain           & e-/ADU & 2.0 & 5.2 & 2.6 \\
 full well      & ke- & 80 & 85 & 85 \\
 read noise$^b$ & e- & 12 / 7  & 17 / 9 & 50 \\
 \hline
 \end{tabular}
 \tablefoot{$^a$ the minimum exposure time is given for the full frame. $^b$ Read noise for the slow mode corresponds to exposures of 2~s (correlated double sampling) and 60~s (up-the-ramp sampling); with the fast mode it is for single uncorrelated reads.}
\end{table}

\subsection{NIX imager}
\label{sec:nix}

\begin{figure}
\centering
\includegraphics[width=8cm]{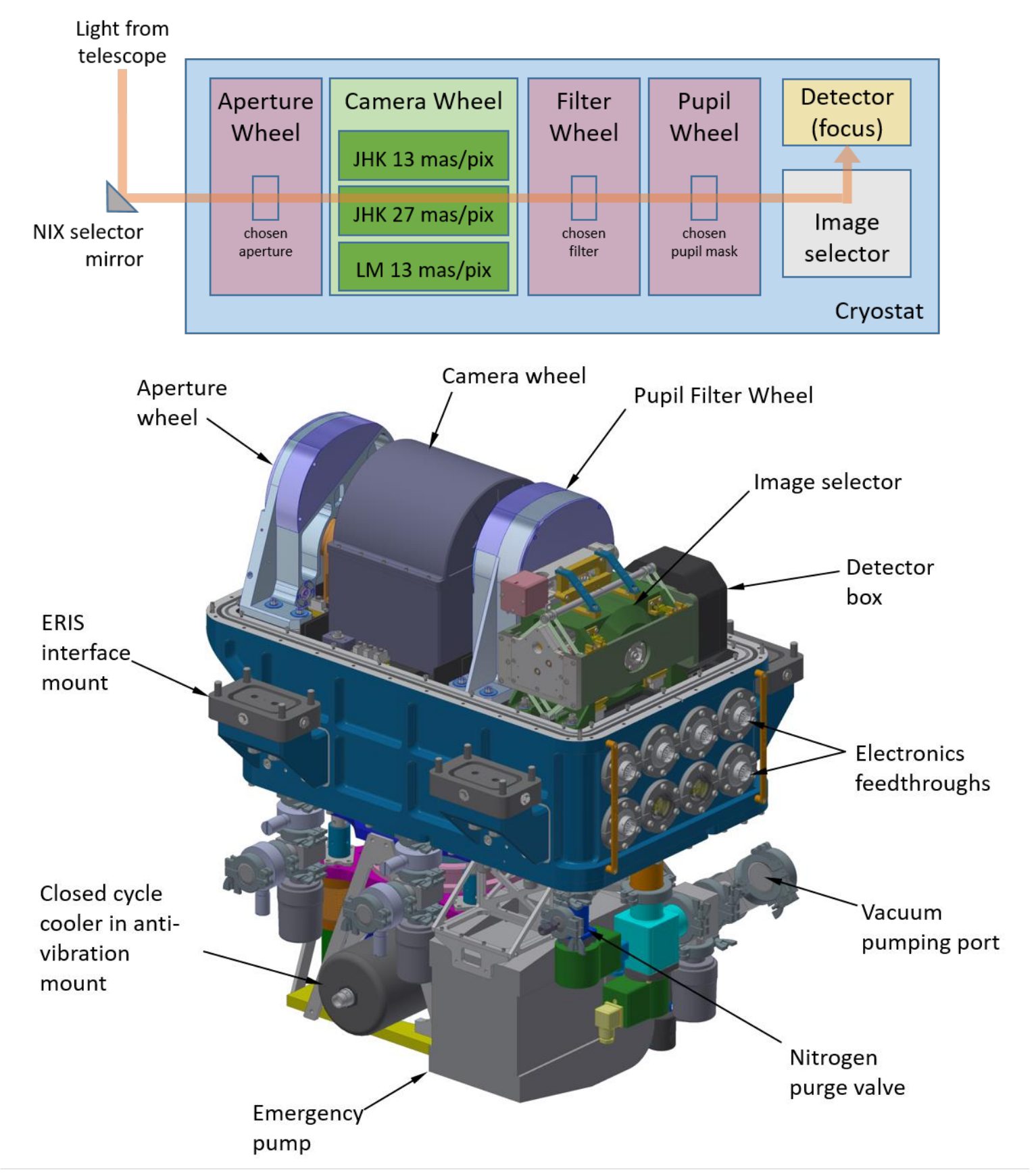}
\caption{Overview of the NIX camera and mechanisms. Top: illustration of the light path through NIX. Bottom: Drawing of NIX with the cover of the cryostat removed so that the cryogenic mechanisms can be seen. The cryostat is cooled with a closed cycle cooler, located underneath in an anti-vibration mount.}
\label{fig:nix}
\end{figure}

\begin{figure}
\centering
\includegraphics[width=\hsize]{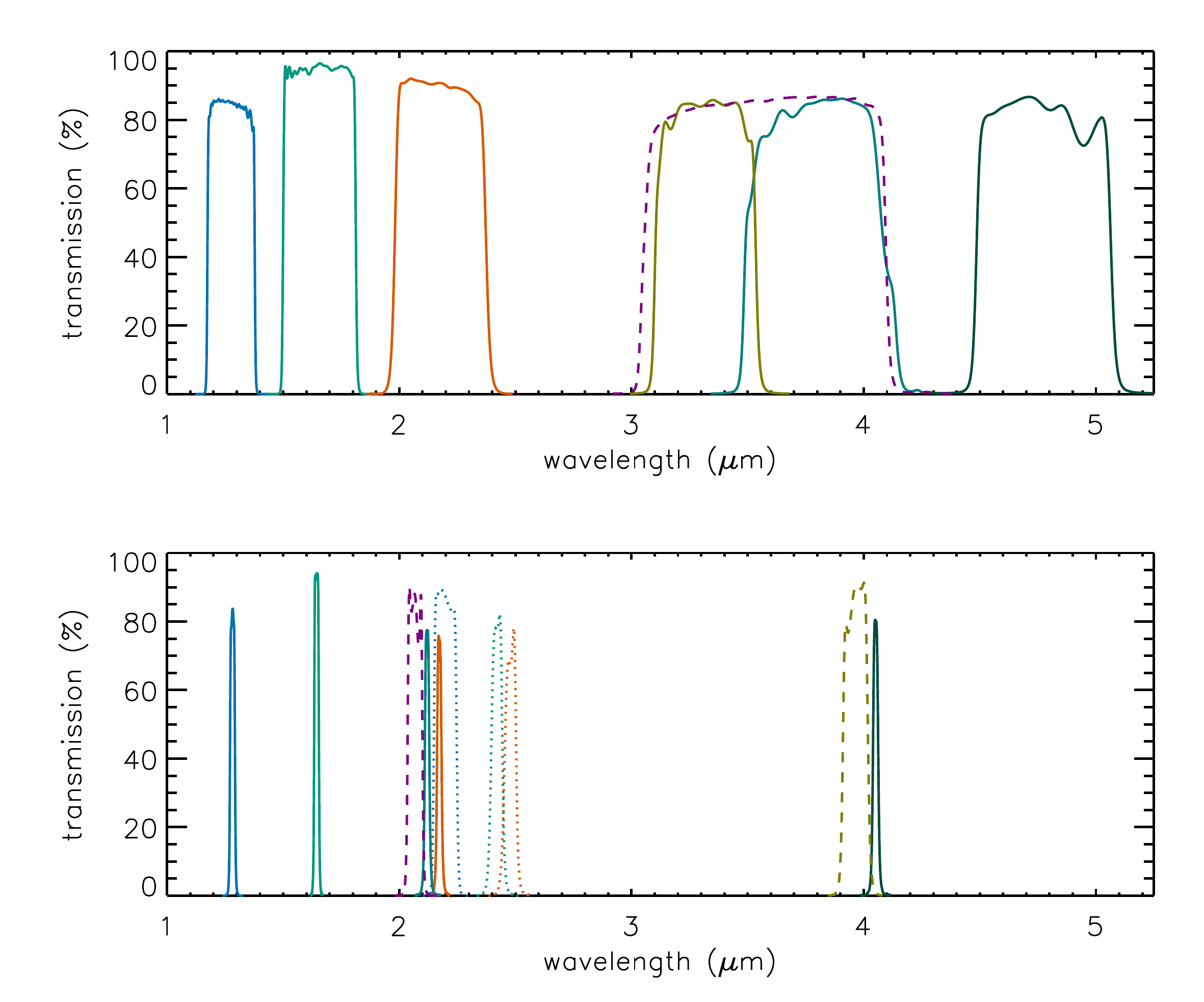}
\caption{Filter profiles in NIX. Top: broad band filters. The dashed profile corresponds to the broad L-band spectroscopic filter. Bottom: narrow band filters. The narrow solid profiles target specific emission lines. Corresponding continuum filters are shown with dashed lines. The other intermediate band filters are shown with dotted lines.}
\label{fig:nixfilters}
\end{figure}

NIX is a compact instrument that weighs only 192~kg and is 725~mm long and 643~mm wide.
It is primarily an imager that provides the capability for diffraction limited imaging at 1--5~$\mu$m, with a variety of high contrast modes and additionally a simple long slit spectroscopic mode \citep{pea16}.
The design is based around three camera assemblies that can each form an image on a Teledyne HAWAII-2RG detector with a 5~$\mu$m cutoff. 
As shown in the top part of Fig.~\ref{fig:nix}, the light enters the cryostat through a CaF$_2$ window before passing through a field stop in the aperture mask mechanism \citep{gla18}, which is set at the telescope focal plane. It then enters the camera barrel where one of three cameras can be selected, covering 1--2.5~$\mu$m at 13~mas and 27~mas pixel scales providing 26\farcs4 and 55\farcs4 fields of view, or 3--5~$\mu$m at 13~mas over a 26\farcs4 field.
Each camera on the wheel comprises three lenses, which are made from BaF$_2$, ZnSe, or IRG2 (a proprietry glass from Schott) and, with only one exception, have spherical surfaces.
After the camera, the light passes through the filter wheel and pupil wheel, which are mounted in a single housing \citep{gla18}.
The filter wheel contains 17 filters (see Fig.~\ref{fig:nixfilters}), as well as open and block positions and one spare slot.
The pupil wheel contains various cold stops, as well as some positions with additional blocking or neutral density filters, and one that includes a grism.
The last mechanism before the detector is the image selector, which has a deployable fold mirror to reduce the path length for the 27~mas pixel scale camera. For the 13~mas pixel scale, the longer back focal length is accommodated using two other fixed fold mirrors.
The detector itself is also on a focus stage. This was needed for laboratory testing, but it is left in a static configuration for all observing modes and small focus offsets between instrument configurations are compensated by the AO system.
To enable pupil imaging, a pair of ZnSe lenses can be inserted into the optical path just in front of the fold mirror of the image selector.
Each of the mechanisms above can be installed, adjusted, or removed without disturbing any of the others.
The total transmission of NIX -- including telescope, atmosphere, and detector quantum efficiency -- for ERIS/NIX has been measured from observations of the spectrophotometric star EG~274 to be 42\%, 61\%, and 52\% in the J, H, and K bands respectively. This is consistent with the estimated throughput of the instrument alone which is in the range 70-75\% for the three bands. The corresponding instrument transmission estimated for the L and M bands is 65-70\%.

\begin{figure}
\centering
\includegraphics[width=9cm]{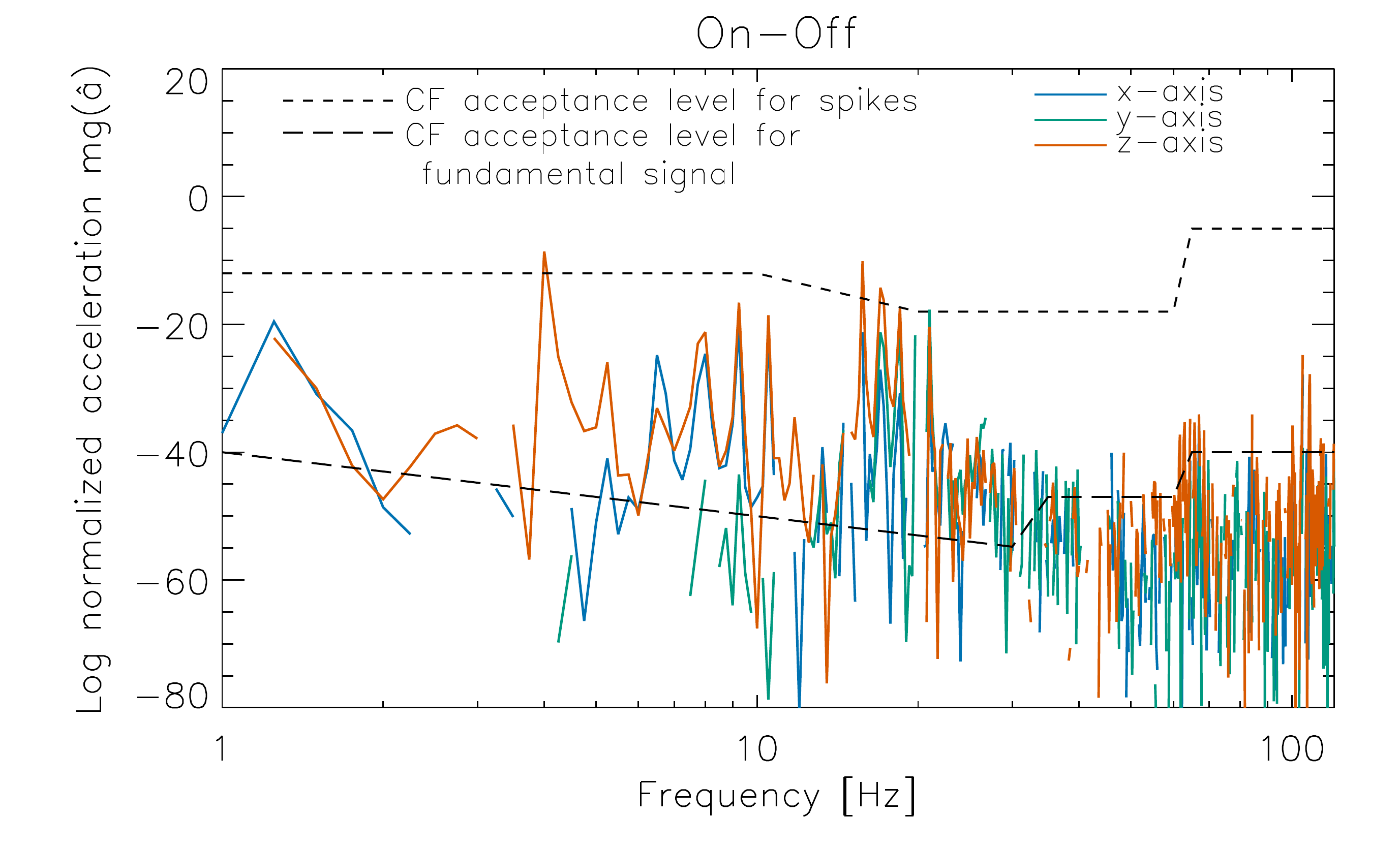}
\caption{Vibrations measured on the attachment flange of ERIS while suspended, shown as a power spectrum of the normalised acceleration (i.e. force, the product of acceleration measured on the flange and adopted mass) after subtracting a baseline measurement. 
The two lines show the requirement for the continuous fundamental signal and that acceptable for spikes in the distribution. ERIS is very close to the specification when suspended in a vertical orientation and if one assumes that the full mass (including support) of 2523~kg is vibrating as a solid body. The vibration decreases when the instrument is inclined by 25$\degr$ in any direction, or if less than the full mass is actually contributing to the force derived on the flange.}
\label{fig:vibration}
\end{figure}

To ensure the instrumental background is low enough, a Leybold 10MD closed cycle cooler keeps the detector assembly at a baseline temperature of 35~K and the other optics at 70~K.
A vibration isolation system, that was developed for KMOS \citep{sha13} is implemented also for NIX. Following a number of iterations and tuning of the spring configuration, the vibration level measured in the laboratory was sufficiently close to the requirement \citep{jak14}, as shown in Fig.~\ref{fig:vibration}. After installation on the telescope, and following adjustment to properly secure the transport lock open, the vibration level transmitted to the telescope has not led to any reports about adverse impact on other VLT, and more importantly VLTI, instruments.

In order to work optimally with the very different background levels in the various bands, the detector has a fast and a slow read mode, the characteristics of which are reported in Table~\ref{tab:detectors}.
The slow read mode is used at 1--2.5~$\mu$m, in a similar way as is done in SPIFFIER, although the read noise is higher because different pre-amplifiers are used.
The fast read mode is used for L and M bands, and enables the full frame to be read out fast enough that the background itself does not saturate the detector.
Additionally, the number of rows read can be reduced, providing up to a factor 4 further increase in the frame rate.
To achieve these rates, the detector is read out in single uncorrelated mode with a correspondingly high read noise.
However the background noise is so significant that it will always dominate in broad-band exposures.
Calculations indicate that in L-band the background from sky contributes typically 55\% of the total background, with another 25\% from the telescope, and ERIS (including the Warm Optics, as well as scattered light) making up only 20\%.
For a 10~ms exposure, the detector noise has an equivalent contribution at the level of around 5\%.
In M-band, the sky is brighter still, contributing nearly 90\% of the total background.

\begin{figure}
\centering
\includegraphics[width=7cm]{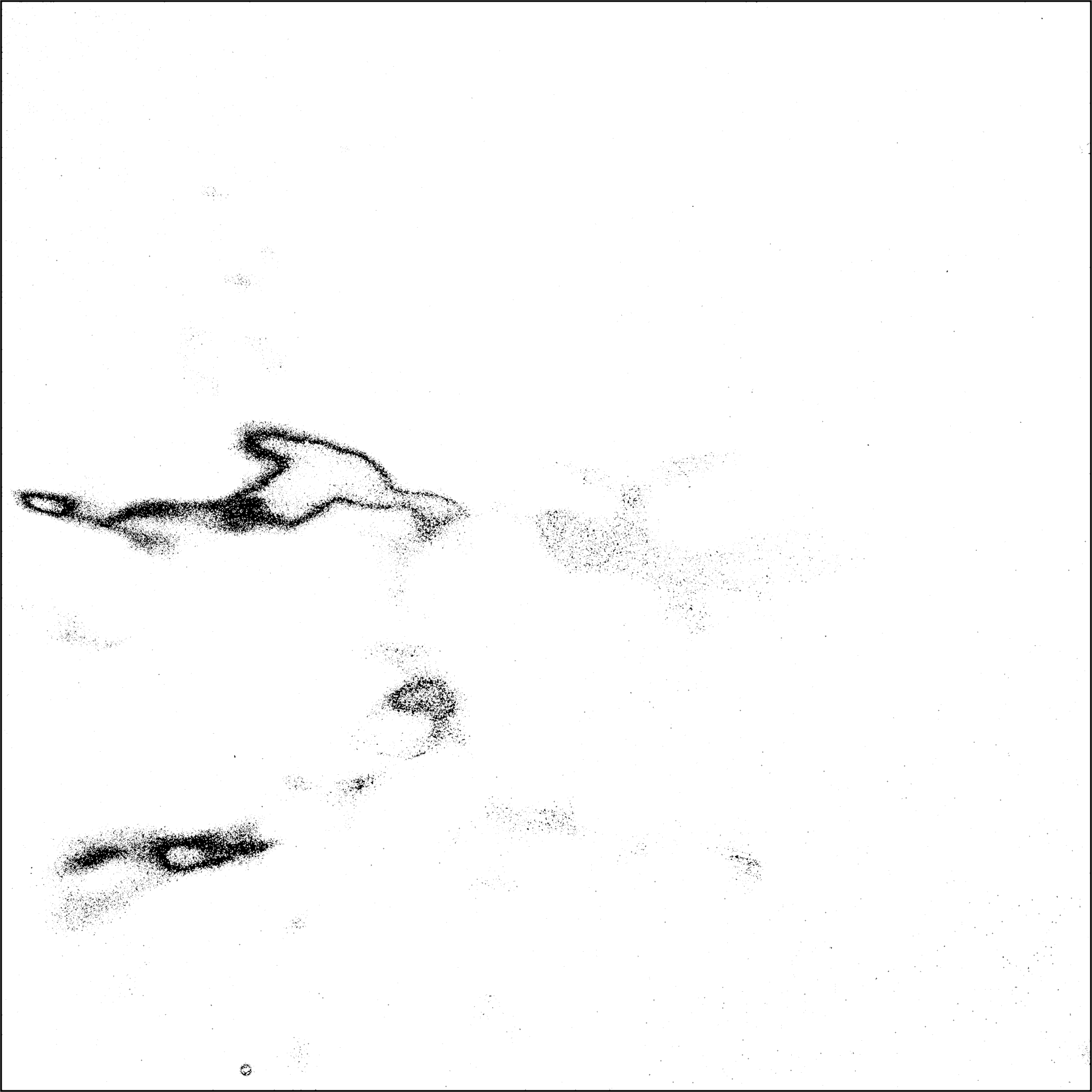}
\caption{Image of the bad pixels in the NIX detector that are adjacent to other bad pixels, on a greyscale from 0 to 8. This representation highlights regions of clustered bad pixels, while ignoring lone ones that can easily be corrected. To indicate the edge of the frame, the four rows/columns of reference pixels all around the edge of the $2048\times2048$ detector have also been marked as bad pixels.}
\label{fig:nixbadpix}
\end{figure}

Unfortunately, the cosmetic quality of the NIX detector is not optimal, with nearly $3$\% bad (unusable) pixels. As shown in Fig.~\ref{fig:nixbadpix}, there are two large areas where many bad pixels are contiguous. However, Sec.~\ref{sec:niximaging} indicates that the impact of this on the final data can be mitigated via sufficient dithers or suitable choice of source position and offsets.

\begin{figure*}
\centering
\includegraphics[width=\hsize]{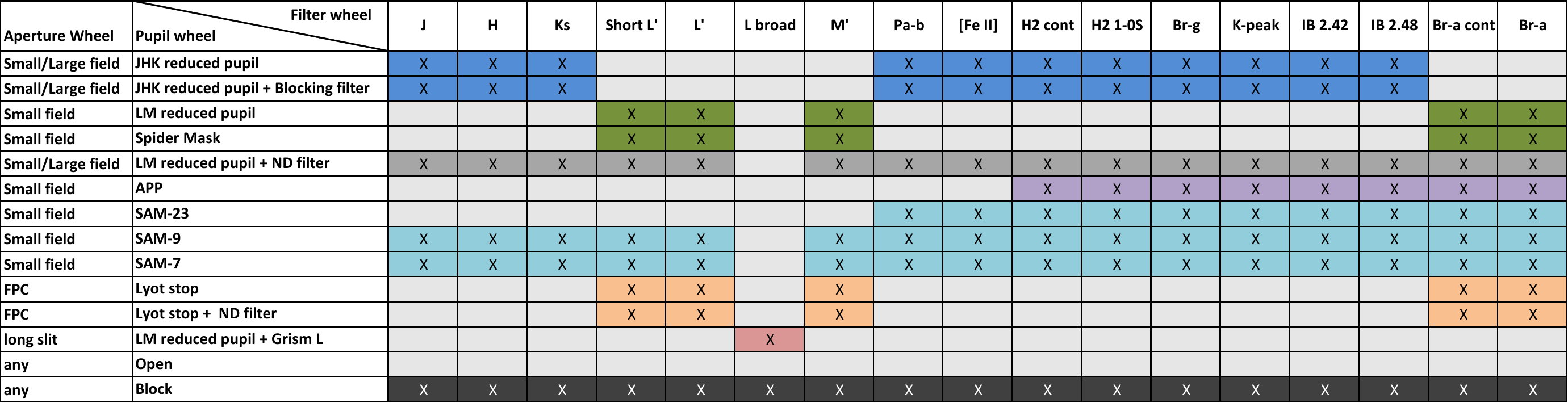}
\caption{Association matrix for NIX, where crosses indicate the nominal combinations of filters and stops allowed (these may differ slightly from those offered in a given observing period). The colours denote observing modes: short wavelength imaging (dark blue), long wavelength imaging (dark green), imaging with a neutral density filter (grey), focal plane coronagraphy (mauve), sparse aperture masking (light blue), focal plane coronagraphy (orange), longslit spectroscopy (red), and a closed light path (black). Other combinations would have to be set manually from the engineering panel.}
\label{fig:nixassoc}
\end{figure*}

In addition to standard imaging, NIX includes a variety of other configuration options.
These are apparent from the association matrix in Fig.~\ref{fig:nixassoc} which indicates which filters can be used for each pupil stop/mask, as well as which aperture wheel positions are available in each case.
The first five rows refer to the imaging configurations, the next five to the high contrast modes, and one more for spectroscopy:

\begin{itemize}

\item
Short wavelength (1--2.5~$\mu$m) imaging is offered with the small or large field of view. A filter that provides additional blocking at wavelengths $>2.6~\mu$m is available. Laboratory tests indicated that this might have been needed for some filters, but observations on-sky have not confirmed measurable long wavelength leakage. 

\item 
Long wavelength (3--5~$\mu$m) imaging is available with the small field. The default cold stop is a metal plate with an undersized circular aperture, which is used with field tracking. A spider mask that blocks background from the central obscuration and secondary mirror supports, is included to minimize the background when imaging in pupil tracking mode.

\item
Imaging with a neutral density filter is an additional option for very bright sources, and is enabled for the small field at short and long wavelengths.

\item
\begin{figure}
\centering
\includegraphics[width=8cm]{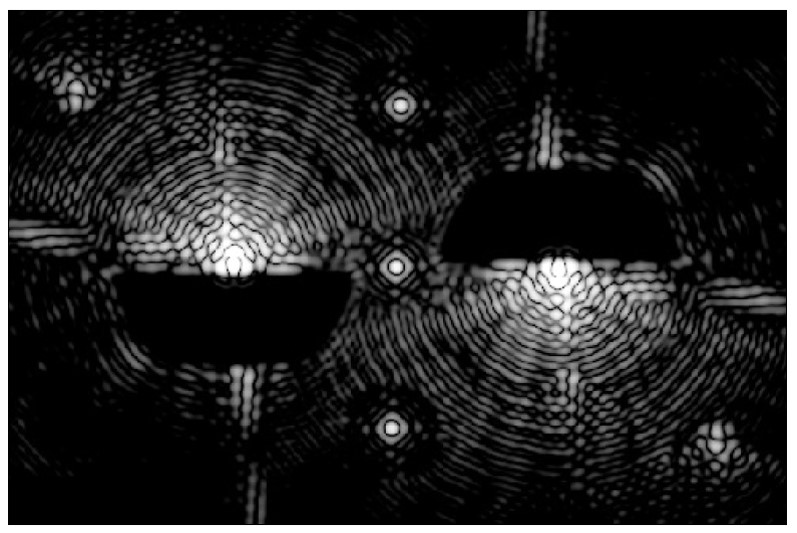}
\caption{The monochromatic theoretical PSF of the grating vector gvAPP used in ERIS, taken from \citet{boe21}. It produces two coronagraphic images, each with a D-shaped dark hole from 2 to 15 $\lambda$/D. A third non-coronagraphic PSF 2\% leakage term is generated for astrometry and photometry, as well as other spots for astrometric reference.}
\label{fig:nixapp}
\end{figure}
A grating vector apodized phase plate \citep[gvAPP; ][]{ott14,doe21} mounted in the pupil wheel provides coronagraphic suppression for all point sources in the field. For each star, the gvAPP in NIX generates two coronagraphic images of the star with dark D-shaped regions on opposite sides, each with 10-30\% of the flux depending on the filter. The two dark holes cover a radial range of about 2-15~$\lambda/D$, and nearly 360\degr\ when combined. A third 2\% leakage term provides a unsaturated non-coronagraphic PSF image of the star that can be used for precise photometry and astrometry, as shown in Fig.~\ref{fig:nixapp}. The gvAPP is used with pupil tracking, and allows for dithering and beam switching on the detector with no loss in science observing time. Because the separation of the two APP PSFs scales linearly with wavelength, the APP is best used with narrow band filters.

\item
\begin{figure}
\centering
\includegraphics[width=8cm]{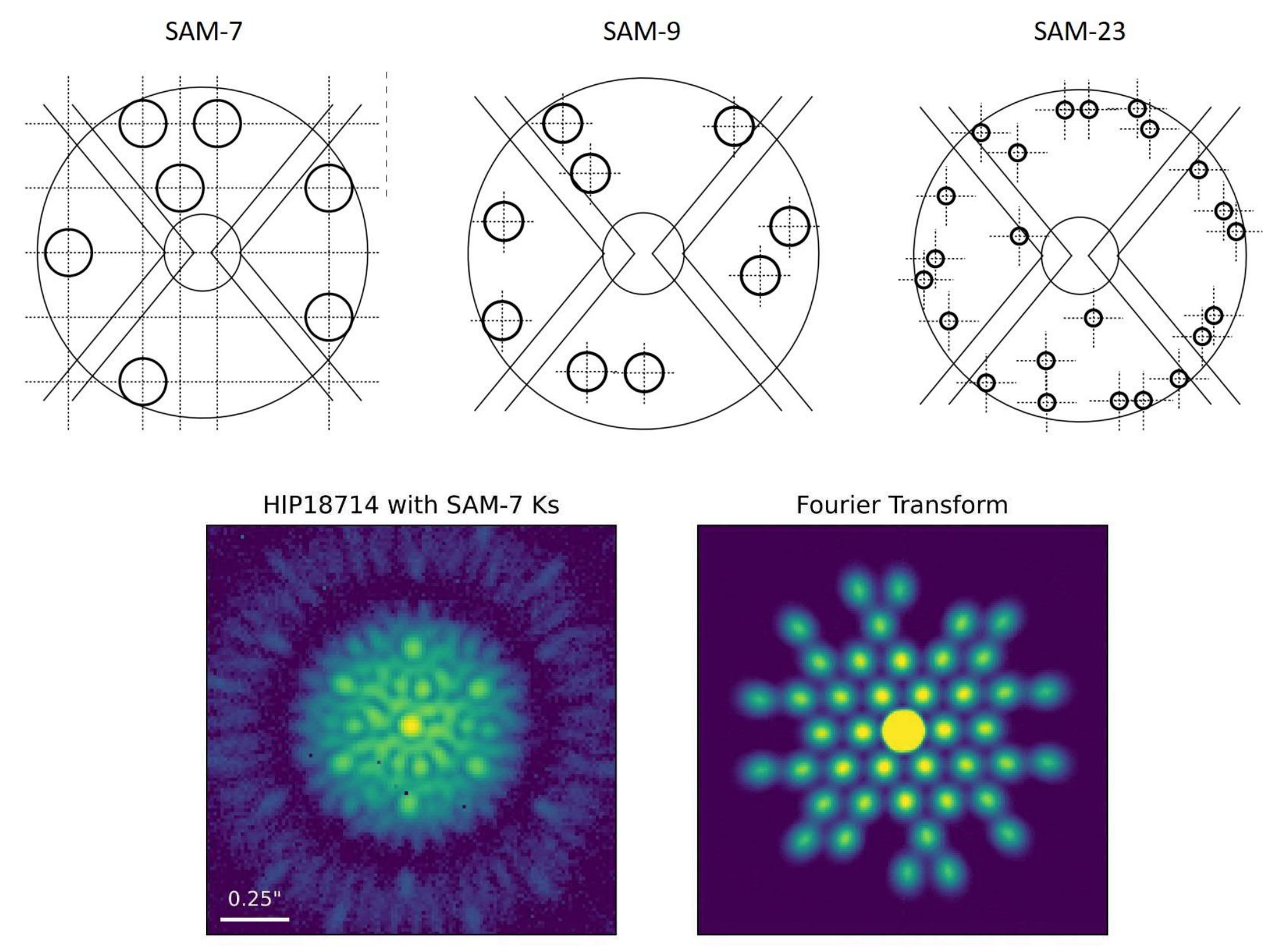}
\caption{The sparse aperture masks in NIX. Top row: drawings of the masks, showing their apertures. Bottom row: The image obtained through SAM-7 (showing the interferogram caused by combination of the non-redundant apertures within the Airy disk of a single aperture), and its Fourier transform.}
\label{fig:nixsam}
\end{figure}
Three sparse aperture masks (SAMs) in the pupil wheel provide a second high contrast mode. They transmit the light through the specified number of mini-apertures for both non-redundant and partially redundant interferometric imaging as illustrated in Fig.~\ref{fig:nixsam}. The mask with seven apertures, providing 21 baselines, is designed to be used on faint targets; the mask with nine apertures, and therefore 36 baselines, is suited to more complex systems; and the mask with 23 apertures, and thus 253 baselines of which 173 are non-redundant, works best on bright targets and yields a resolution a factor two better than the diffraction limit of the telescope. Their throughputs, relative to the JHK and LM imaging pupil stops, are in the ranges 12-15\%, 9-11\%, and 5-6\% respectively.

\item
\begin{figure}
\centering
\includegraphics[width=5.8cm]{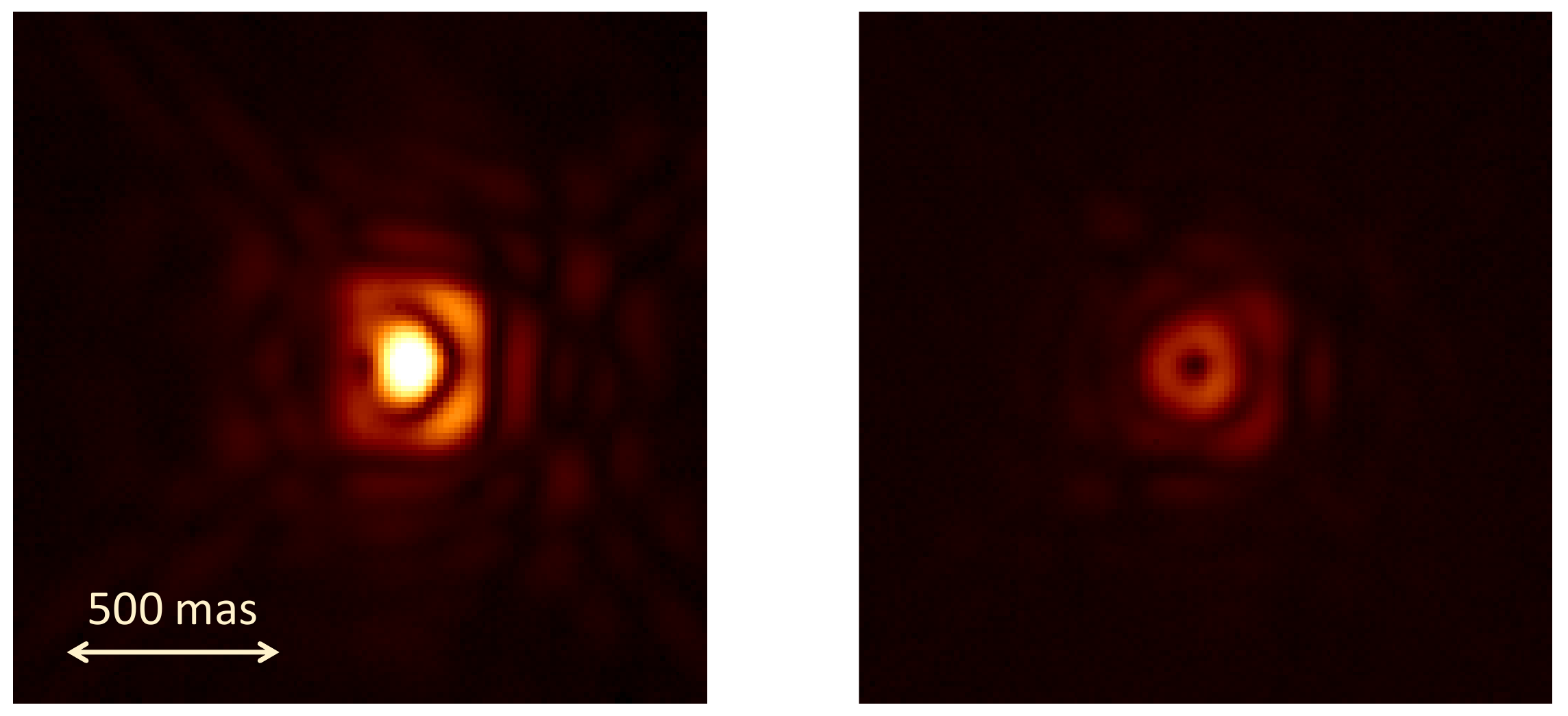}
\caption{Improvement of the suppression of the on-axis star with better centering of the FPC in NIX (images are displayed with a square-root stretch). The left image has an offset between the star and the coronagraph of about 1~$\lambda$/D (95~mas), at which point the suppression is about a factor 2. The right image shows a well centered image, where the rejection of the stellar light is more than a factor 100. The scale is shown; for reference the FWHM in L$^\prime$ band is 97~mas and the first Airy ring has a diameter of 310~mas.}
\label{fig:nixfpc}
\end{figure}
Focal plane coronagraphy (FPC) is implemented using an annular groove (or vortex) phase mask \citep{maw05}, followed by a Lyot stop in the pupil plane. There are two masks, one for each of the L and M bands. The combined throughput (vortex phase mask and Lyot stop) amount to $\sim$62\% in L band and $\sim$51\% in M band. The vortex coronagraph is one of the most efficient in terms of inner working angle, with a discovery space starting already at $\sim$1~$\lambda$/D, and works well with broad bands. To perform optimally, the star needs to be precisely centered on the mask, in which case it suppresses the central PSF significantly, leaving only a ring of residual light, as shown in Fig.~\ref{fig:nixfpc}. The star and mask are kept aligned during the observation using an automated algorithm called Quadrant Analysis of Coronagraphic Images for Tip-Tilt Sensing (QACITS; \citealt{hub15}) that provides an offset correction every few seconds.

\item
Long slit spectroscopy (LSS) is possible over the 3.05--4.05~$\mu$m range at a spectral resolution of $R\sim900$. It requires the $12\arcsec\times86$~mas slit in the aperture wheel to be used with the \code{L-broad} filter and a grism next to the cold stop in the pupil wheel. Standard nodding techniques are used during operation. The calibration of the spectral distortion and wavelength solution are considered static, and as such are updated only occasionally.

\end{itemize}

\begin{figure}
\centering
\includegraphics[width=7cm]{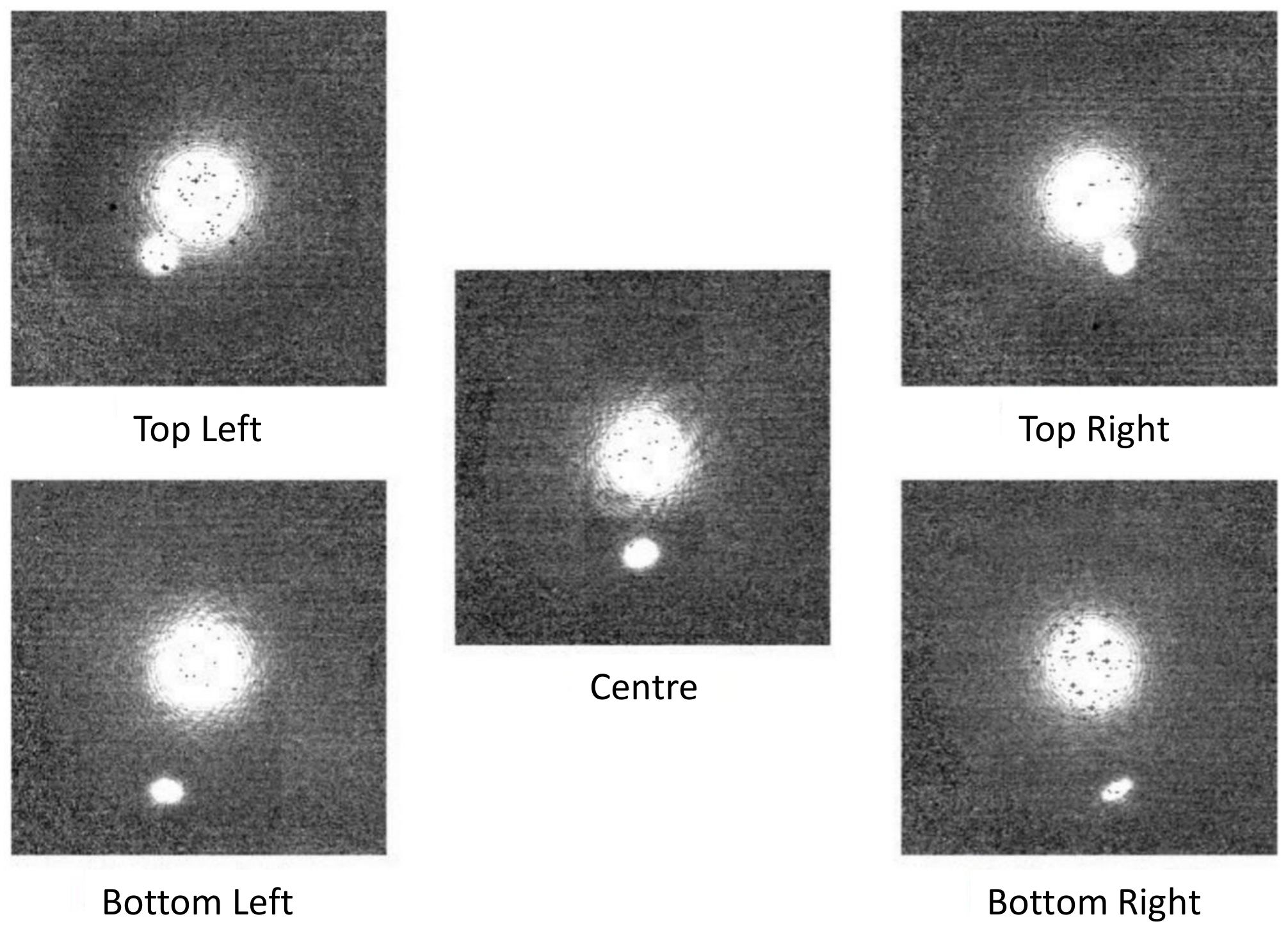}
\caption{Images of the ghost that occurs with the Pa-b filter, showing how it moves with respect to the primary source in a well defined way at different locations across the detector. The scaling of the images is set to a low level to emphasize the ghost, which has a peak intensity only 0.2\% that of the primary source.}
\label{fig:nixghosts}
\end{figure}

Finally, as for all cameras, NIX does have optical ghosts, despite efforts to minimise them.
To mitigate inter-optical ghosts, such as out-of-focus reflections from the detector to the filters and back again, the 8~mm thick CaF$_2$ filters are tilted by 4\degr. 
More difficult to address are the intra-optical reflections, which create faint ghosts that are more in focus. These can be reduced by having very high throughputs for the filters, so that the internally reflected component is minimised.
With typical filter transmissions in NIX of 80--90\%, the ghosts generally have a peak flux ratio of $\sim0.1$\% with respect to the primary PSF; for the narrow band filters with low transmissions in the range 60--70\%, this ratio increases to 0.5--1\%. The ghosts are offset from the primary PSF by 0.3--0.6\arcsec, in slightly different directions below the PSF. The change in relative location of the ghost across the detector is well defined. An example is given in Fig.~\ref{fig:nixghosts}, and for the \code{Br-g} filter one finds the shift of the ghost 
$\delta (x,y)_{ghost}$
with respect to the source is 
$\delta x_{ghost}/\Delta X_{source} = \delta y_{ghost}/\Delta Y_{source} = 0.0258$, where 
$\Delta (X,Y)_{source}$ 
is the shift of the source on the detector.
This means that in a dithered set of images the impact of the ghost in the combined product will be reduced to significantly lower levels than in single exposures.

\section{Main Science Drivers on the Instrument Design}
\label{sec:scidriver}

ERIS is a workhorse instrument optimised for diffraction limited observations of individual targets, and can be used for a wide variety of different science cases. As motivation for the different observing modes it offers, we begin by outlining three priority science topics that were identified during the instrument's design phases. We then use observations based on these in Sec.~\ref{sec:commresults} to illustrate the instrument's capabilities and verify realistic operational regimes that may include unexpected challenges.

\subsection{Galaxy Evolution at High Redshift}

The focus here is on the physical processes driving mass assembly and structural transformations of galaxies at redshifts $z \sim 1 - 3$, which encompass lookback times around the peak of the cosmic star formation rate density 8 to 11 billion years ago.  This is the epoch during which a substantial fraction of the stellar mass of local massive disk and elliptical galaxies formed \citep{mad14}.  As much as 90\% of the star formation occurred in galaxies following a tight relationship in stellar mass versus star formation rate.  These "main sequence" galaxies were 5-10 times more gas-rich than their $z \sim 0$ counterparts, hosted 10-20 times higher star formation activity, and consisted predominantly of rotating, turbulent disks \citep{for20,tac20}.  Under these conditions, internal processes such as gravitational instabilities and efficient inflows promoted rapid growth of bulges and central black holes, and outflows driven by star formation and AGN were both energetic and frequent.
The signatures of these processes can be seen most clearly by their imprint on resolved galaxy properties: given the typical gas fraction and turbulence of high redshift disks, they occur spatially on characteristic scales of $\sim 1$~kpc (the Toomre scale, corresponding to $0\farcs 1$ at $z \sim 1 - 3$) and kinematically at levels as low as $\sim 10-50$~km\,s$^{-1}$ \citep[e.g.,][]{bour07,gen11,gen20,cev12,price21}.
Enhanced turbulence and gas transport should further contribute to efficient mixing of metals within galaxies, alongside infall of pristine gas and ejection of enriched material via outflows, thus leading to flattened metallicity gradients \citep[e.g.][]{cresci10,wuyts16,curti20,wang20}.

Observational constraints of the relevant properties on the relevant scales are scarce.  Indeed, the number of distant galaxies that have been observed with adaptive optics-assisted integral field spectroscopy (AO-IFS) is limited, with $\sim$\,200 at $1 < z < 3$ \citep{for20}.  The largest coherent samples are modest in size, such as the SINF/zC-SINF AO survey with 35 objects at $z \sim 1.5 - 2.5$ \citep{for18}, or the sHiZELs sample of 34 sources in narrow $z = 0.84$, 1.47, 2.23, 3.33 intervals \citep{gillman19}. Overall, the AO-IFS samples had diverse primary selection criteria and represent a heterogeneous collection probing different parts of mass, star formation rate, and redshift space, with increasing bias towards higher-than-typical star formation activity at lower masses.
In contrast, seeing-limited IFS samples comprise 10 times as many objects ($\sim 2000$) and probe better the typical main sequence star-forming population at $z \sim 1 - 3$ and stellar masses $M_{\star} \ga 10^{10} M_{\odot}$, albeit with coarser resolution of $\sim 5$~kpc \citep{for20}.  Here, the efficient multi-IFS instrument KMOS at the VLT greatly facilitated the census of many hundreds of such galaxies -- e.g. through the KROSS and KMOS$^{3D}$ surveys with nearly 800 targets each \citep{harr17,wis19}.

ERIS is a key instrument for enlarging the AO-IFS samples, and will go well beyond the determination of kinematics, star formation, metallicity, and outflow properties on global galactic scales.  With scientific requirements aimed at elucidating the physics driving galaxy evolution at $z \sim 1 - 3$, it has capabilities optimized notably to quantify non-circular motions and radial gas transport, characterize individual giant star-forming clumps, detangle outflows driven by star formation across disks and by AGN in the nuclear regions, and map the gas velocity dispersion and metallicity on the relevant spatial and velocity scales.
It will do so far better and more efficiently than SINFONI or any other existing ground-based near-infrared AO integral field spectrograph thanks to greatly improved Strehl ratio, spectral resolution, and sensitivity.  With high spectral resolution of $R$\,$\sim$\,5000 and $\sim$\,10000, it will outperform JWST in terms of kinematics, for which spectroscopic modes are limited to $R \la 2700$.
With excellent AO performance and SPIFFIER's high resolution gratings, observations using ERIS will play a key role in probing the physical mechanisms of galaxy evolution at the heyday of massive galaxy formation.

\subsection{Direct Imaging of Exoplanets}

The 3–5~$\mu$m wavelength regime, covering the L and M bands, plays a central role in the discovery and characterisation of exoplanets through their thermal emission \citep[e.g. see the reviews of ][]{curr22,Bowler16}.
Indeed, exoplanets are usually much colder than their host star. This leads to the ratio between the planetary and the stellar flux, i.e. the contrast, becoming more favourable at longer wavelengths. 
Theoretical models of gas giant exoplanets show very red colours over ages up to 300 Myr, and the fluxes at shorter wavelengths fade at a faster rate compared to the longer wavelengths - this makes longer wavelength L and M band imaging competitive despite the higher sky thermal background.
This is the reason why, before the advent of dedicated high contrast instruments, most detections of exoplanets were in the L band, for example the discovery of $\beta$~Pictoris~b a 4–11~M$_{Jup}$ planet orbiting at a distance of 9~AU from the star in a plane close to that of the debris disk at larger scales \citep{lag10}.
Following the first L band survey \citep{kas07,kas09}, several surveys in this band have been carried out with first generation high contrast instruments such as NACO and Clio (currently on the Magellan Telescope). They are sensitive to older or lower mass planets which are colder, and it remains the rationale behind the more recent exoplanet survey reported by \citet{sto18}.
Longer wavelengths are also competitive for detecting very young proto-planets that are still highly obscured by circumstellar and circumplanetary dust. 
The warm (few 100~K) circumplanetary material in which they are embedded provides a large emitting area, making coronagraphy and sparse aperture masking in the L and M bands efficient tools to characterise them.
In the case of HD~100546, the protoplanet was detected in both of these bands, but not in K-band \citep{qua13,qua15}.
Finally, the Bracket-$\gamma$ and Bracket-$\alpha$ filters of NIX offer the opportunity to study emission lines created by the ionized gas in accretion shocks of forming exoplanets at wavelengths less subject to extinction than H$\alpha$, which was successfully used to detect a forming proto-planet in the gapped disk of the PDS 70 system \citep{Wagner2018,Haffert2019}.

In addition to detecting very young and old planets, ERIS provides a unique probe of the atmospheres of gas giant planets, particularly in the context of non-equilibrium chemistry and clouds. 
Combining detections in the J, H, and K bands with follow-up observations by ERIS in the L and M bands will vastly improve the efficacy of the scientific analysis by providing a longer baseline for fitting atmosphere models.
For the case of two planets in the HR~8799 system (comprising four gas giant planets of 7–10~M$_{Jup}$) standard cloudless atmosphere models cannot explain the observed photometry, particularly in the L band.
Instead, patchy cloud prescriptions provide the best match to the available data \citep{cur11,ske12}.
Such work is in its infancy, and the sensitivity and 3–5~$\mu$m capabilities of ERIS, focusing on high throughput and a small inner working angle, open the potential for exciting advances that are complementary to other high contrast facilities aiming to directly image exoplanets at optical wavelengths.

The spectrograph SPIFFIER provides a different path to investigating the chemical composition of the atmospheres of gas giant exoplanets. For example, the K band offers the opportunity to study the CO bandhead at 2.3~$\mu$m \citep{martin97}. It will build on the earlier work done with SINFONI which lead, for instance, to the detection of CO and H$_{2}$O in the atmosphere of the young $\beta$ Pic b, to the constraint on the isotopologue ratio $^{13}$CO/$^{12}$CO in the accreting planet TYC 8998 760 1 b, and to the constraint on the atmospheric parameters of HIP 65426 b \citep{Hoeijmakers2018,Zhang2021,Petrus2021}.
With its improved adaptive optics, ERIS is expected to reach deeper sensitivity limits, ultimately allowing the measurement of higher SNR spectra for faint, well-separated low-mass companions. 
However, the number of available high-contrast targets at small angular separation is limited by the brightness of the host star, a restriction that can only be overcome by putting the star outside the field of view, which complicates the image centering step in the post-processing. 

A potential improvement for the future would be the implementation of pupil tracking for SPIFFIER, for which ADI would enable the extraction of the full spectra of low-mass companions on close orbits. 
Currently, the molecular mapping technique \citep{Hoeijmakers2018} yields an alternative way to investigate the chemical composition of such challenging targets without the need for ADI. Indeed, this technique works by removing the continuum of the measured spectra via high-pass filtering, essentially removing the stellar PSF together with the continuum of the planet. Due to the major differences between stellar and exoplanet spectra, the residual light can be cross-correlated with spectral templates of the exoplanet atmosphere to pick up its signal.

\subsection{Galactic Center}

The vast improvements over the last two decades in spatial resolution (to 50~mas, equivalent to 0.002~pc) and sensitivity (by 3-5~mag) made possible by AO on the VLT, have revealed surprises in the Galactic Center and challenge theories of star formation and gas inflow in galaxy nuclei \citep{gen10}.
One of the most high profile results has come about through frequent and regular monitoring of the 3D orbits of the stars closest to Sgr~A* via their proper motions and line-of-sight velocities. 
Of particular importance are stars on elliptical orbits, which can cover a large radial range very rapidly and can show changes in their radial velocity of hundreds to thousands of km\,s$^{-1}$ within timescales of months.
AO observations that track these have provided exquisite constraints on both the mass of and distance to the supermassive black hole \citep{boe16,gil17}. These have been exceeded only by interferometric observations from GRAVITY that achieve unprecendented precision by combining all four 8-m UTs of the VLT \citep{gra19}.
The AO spectroscopic data retain a central role in terms of the precision with which one can test general relativistic effects.
This is most obvious in the detection of the gravitational redshift of the star S2, which is essentially a velocity measurement \citep{gra18,dot19};
but for which the proper motions of the stars are also needed to provide the necessary constraints on the orbital model.
Similarly, the detection of the relativistic precesssion of S2 is essentially a proper motion measurement \citep{gra20}; but the spectroscopic data -- and their uncertainties -- are needed to constrain the orbital model and hence contribute to the measurement.

Other discoveries that have come about via astrometric imaging and integral field spectroscopy -- capabilities that lie at the core of ERIS -- include the differing spatial distributions of late type stars, early type stars, and Wolf-Rayet stars, as well as the various 3D orbital structures they trace. 
The discovery of a gas cloud G2 falling towards Sgr~A* \citep{gil12} generated considerable interest in the community, as well as numerous theories about what it is and where it has come from. While visible at L band, it has not been detected at K band, putting strong constraints on the dust temperature \citep{phi13}. 
It is very clearly visible in recombination lines such as Br$\gamma$, allowing the change in velocity shear over time to be tracked. That this can be matched by a simple set of test particles moving in the black holes potential suggests that the ionised gas is not bound together \citep{pfu15}.

\begin{figure*}
\centering
\includegraphics[width=15cm]{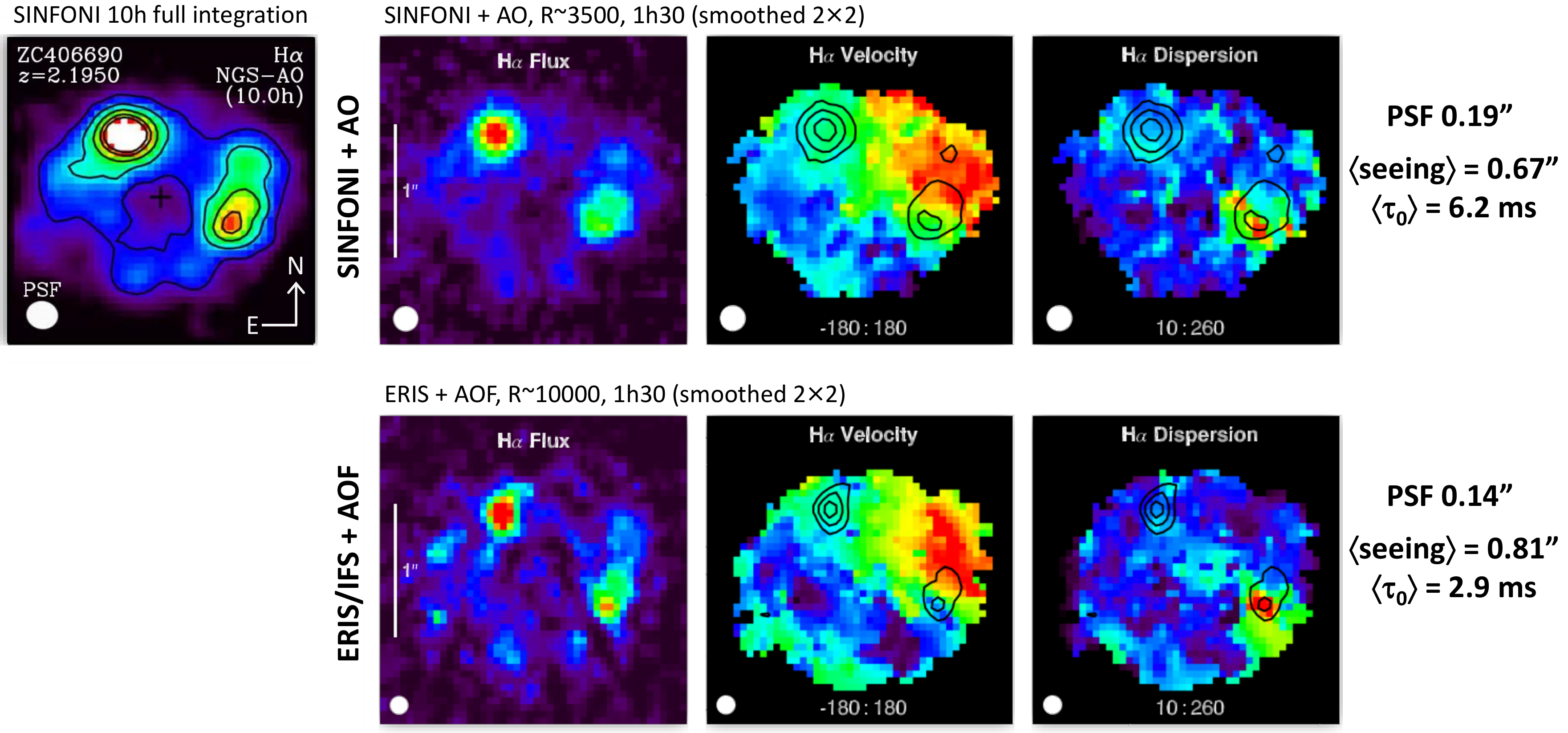}
\caption{Flux and kinematics maps of the $z=2.19$ galaxy ZC406690, comparing LGS-AO integral field spectroscopy between SINFONI (top) and ERIS (bottom). In both cases the data have been lightly smoothed. Far left: the full 10~hr archival SINFONI flux map. Top: a 90~min subset of those data (showing flux, velocity, and dispersion), taken in reasonably good conditions. Bottom: the 90~min ERIS dataset taken at $R\sim10000$ in less optimal conditions. ERIS data were taken early in commissioning before the AO system was optimised, but already out-perform the equivalent $R\sim3500$ SINFONI dataset, and show all the structure visible in the full SINFONI integration.}
\label{fig:galev}
\end{figure*}

In the Galactic Center, the key science topics include dynamics of the various stellar populations, tracking and characterisation of infalling gas clouds, the radiative behaviour of Sgr~A* via its flares, and continued monitoring of fainter and closer stars around Sgr A*. The instrumental requirements to address these are among those for which ERIS is optimised: H to L band astrometric imaging together with H and K band integral field spectroscopy. In all cases a critical criterion is resolution, for which large mirror size and the ability to observe at H band (at shorter wavelengths extinction is prohibitive) are both important. 

\section{Early Results from Commissioning}
\label{sec:commresults}

A number of the commissioning tasks were performed on science targets. The rationale for doing so includes testing every aspect of the instrument on the types of targets for which it will ultimately be used. Because this involves the science team in preparing the templates that are used for the observations, it leads to novel configurations and sequences that would not necessarily have been tried out in other circumstances. Similarly, it allows the data processing pipeline to be tested in realistic situations.
Additional benefits are that one can use external data to quantify some ERIS parameters, or verify the fidelity of ERIS data by comparison to existing data.
While the astrophysical rationale is strongly present for these observations, we focus here on technical issues rather than providing detailed scientific analyses of the targets observed.
We also do not provide details about sensitivities since these are strongly dependent on the ambient atmospheric conditions and AO performance, which is itself dependent on the magnitude and offset of the guide or tip-tilt star. Instead, at the end of the paper we provide a link to the relevant website where potential users can assess the expected performance for their specific use case.

\subsection{Integral Field Spectroscopy}

\paragraph{ZC406690}

During one of the early commissioning runs ERIS was pointed towards ZC406690 at $z=2.195$, for which the H$\alpha$ line is redshifted into the K band at 2.097~$\mu$m. It is a $M_* = 4\times10^{10}$\,M$_\odot$ star forming galaxy close to the main sequence with a total $H\alpha$ flux of $3\times10^{-16}$~erg\,s$^{-1}$\,cm$^{-2}$ that had previously been targeted with SINFONI \citep{for18}. The line emission in the disk of the galaxy is distributed in a star forming ring, broken up into several prominent clumps, rather than concentrated in the centre. Observations were performed using SPIFFIER with LGS-AO and a tip-tilt star of $G_{RP} = 14.7$~mag that was 19\arcsec\ away. Since the target cannot be detected in the short exposures used for acqusition, it provides the opportunity to test some additional operational procedures, specifically a blind offset from a reference star -- in this case the tip-tilt star -- which is also used as a PSF reference. The observations include a series of dithered 300~sec exposures interspersed with offsets to sky. In order to include the full disk of the galaxy within the field of view, the intermediate pixel scale was used, which precludes reaching the diffraction limit. However, this early during the commissioning, the AO performance was not optimised. The aim was instead to perform initial tests of stability over long integrations, offsets to sky, the PSF calibration template, and assess the spectral flexure of the new high resolution grating in SPIFFIER.
The spatial resolution achieved during the 90~min on-source time was 0\farcs14 in average conditions, during which the seeing and coherence time (both at 500~nm) were 0\farcs81 and 2.9~ms respectively. In comparison, the resolution for an equivalent set of archival SINFONI observations of the same object was only 0\farcs19 in better seeing and coherence time of 0\farcs67 and 6.2~ms respectively. Even at this early stage, Figure~\ref{fig:galev} shows that the ERIS data provided better definition of the clumps in the galaxy disk, which match up very well with those in the full 10~hr SINFONI dataset.
This is particularly important here because the bright clump on the south east side of the ring shows clear evidence for an outflow (high velocity dispersion), and the ERIS data are better able to localize where this is occurring through the spatially resolved velocity dispersion of the emission line.  These data demonstrate the ability of ERIS to characterize in more detail the size, star formation, local outflows, and other properties of clumps, as key tracers of gravitational instabilities in high-redshift gas-rich turbulent disks.


\begin{figure*}
\centering
\includegraphics[width=\hsize]{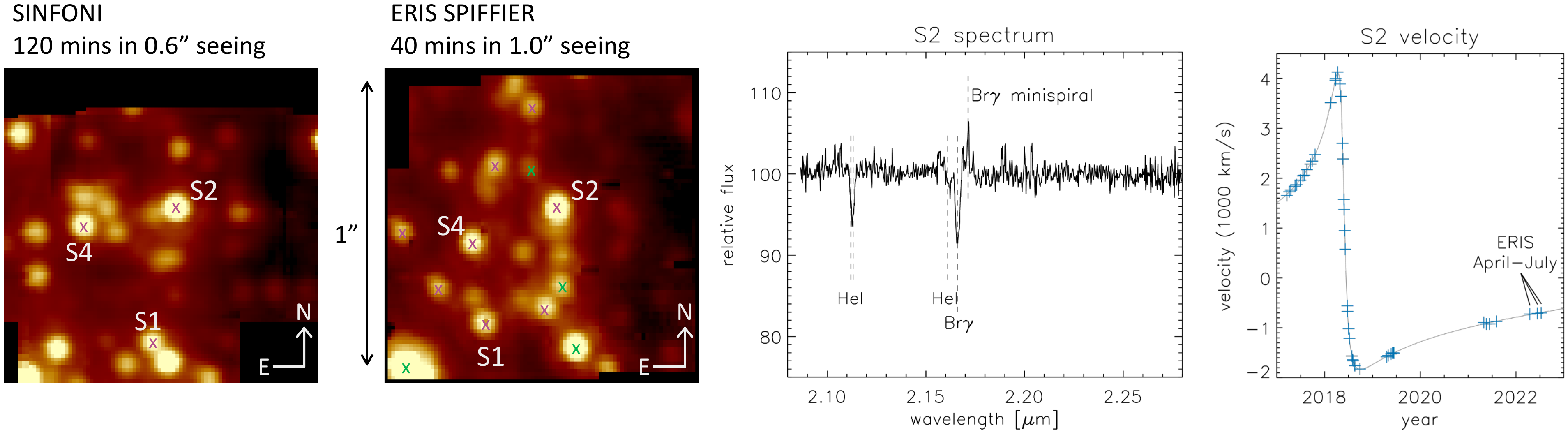}
\caption{Comparison of LGS-AO integral field spectroscopy of the Galactic Center between ERIS (June 2022) and SINFONI (April 2008). The quality of the 40~min ERIS data (centre left), which were taken in 1\farcs0 seeing, are comparable to the very best 210~min SINFONI data taken in 0\farcs6 seeing, as evidenced by the diffraction rings that can be seen around S2 and S4. This relatively short integration was sufficient to show clear detections of the He\,I and Br$\gamma$ absorption lines in S2 (centre right), and hence to measure a precise velocity that matches well that predicted from the model of its orbit (far right; adapted and updated from \citealt{gra19}).}
\label{fig:gcspiffier}
\end{figure*}

\paragraph{Galactic Center}

Integral field spectroscopy of the Galactic Center also poses some operational challenges because the field is very crowded and the stars move relative to each.
As such, for each epoch, the field that one sees with the smallest pixel scale of SPIFFIER looks different.
This means that the acquisition has to be done carefully and precisely, a process that provides verification of the slightly different centres of the fields covered by the three pixel scales. In addition the sky exposures require an offset of nearly 14~arcmin to reach a sufficiently empty field. And after remaining at the sky position for 600~sec, a time equal to each on-source exposure, the offset back must return to the same position with a precision of about 1\arcsec\ in order for the AO loop to re-close automatically. Achieving this robustly requires switching back and forth from AO guiding during the object exposures to telescope guiding (rather than just tracking) during the sky exposures.
The results from these tests, performed using LGS-AO with a $G_{RP}$=13.4 star 19\arcsec\ away about halfway through the commissioning, are shown in Fig.~\ref{fig:gcspiffier}, with an image from SINFONI (far left) for comparison.
The image from SPIFFIER (centre left), reconstructed from 40~min of on-source exposure, has several of the known younger and late type stars marked with purple and green crosses respectively; and some of the stars are labelled.
Although these data were taken in moderate seeing of about 1\arcsec, diffraction rings can be seen around the brighter more isolated stars in the field such as those labelled, indicating the high quality AO correction. These were only seen in the best SINFONI data taken in very good seeing.

Measurement of the positions of seven relatively bright and more isolated stars in the field verified that the pixel scale is precise to better than 1\%. Although SINFONI will be affected by the anamorphism induced by the tilted dichroic as described in Sec.~\ref{sec:spiffier}, it corresponds to $<0.1$~pixel at the edge of the field and so is negligible (unlike for NIX, see Sec.~\ref{sec:niximaging}). The orientation of the field is rotated by $\sim1.5$\degr anti-clockwise with respect to the given position angle. This corresponds to half a pixel at the edge of the field, and so is close to the uncertainty with which it could be measured during commissioning.

The orbits of the stars around the supermassive black hole (located just below S2 in the image) have been traced previously using proper motions and line-of-sight velocities.
The spectrum from S2 (centre right) clearly shows the known He\,I and Br$\gamma$ absorption \citep{hab17}. 
The velocity of this star has been monitored over nearly two complete orbits, and the wavelength of the lines can change rapidly when it is close to the black hole. 
The measurements lie at the expected locations in the plot of velocity versus time (far right), in all three cases deviating from the predicted values by less than 10~km\,s$^{-1}$ -- satisfactory given their preliminary status.
In the same small data set, the velocities of more than six other S-stars (including S1 and S4, labelled in the figure) could be measured with sufficient precision to improve the constraints on their orbits.

\subsection{Imaging}
\label{sec:niximaging}

\begin{figure*}
\centering
\includegraphics[width=14cm]{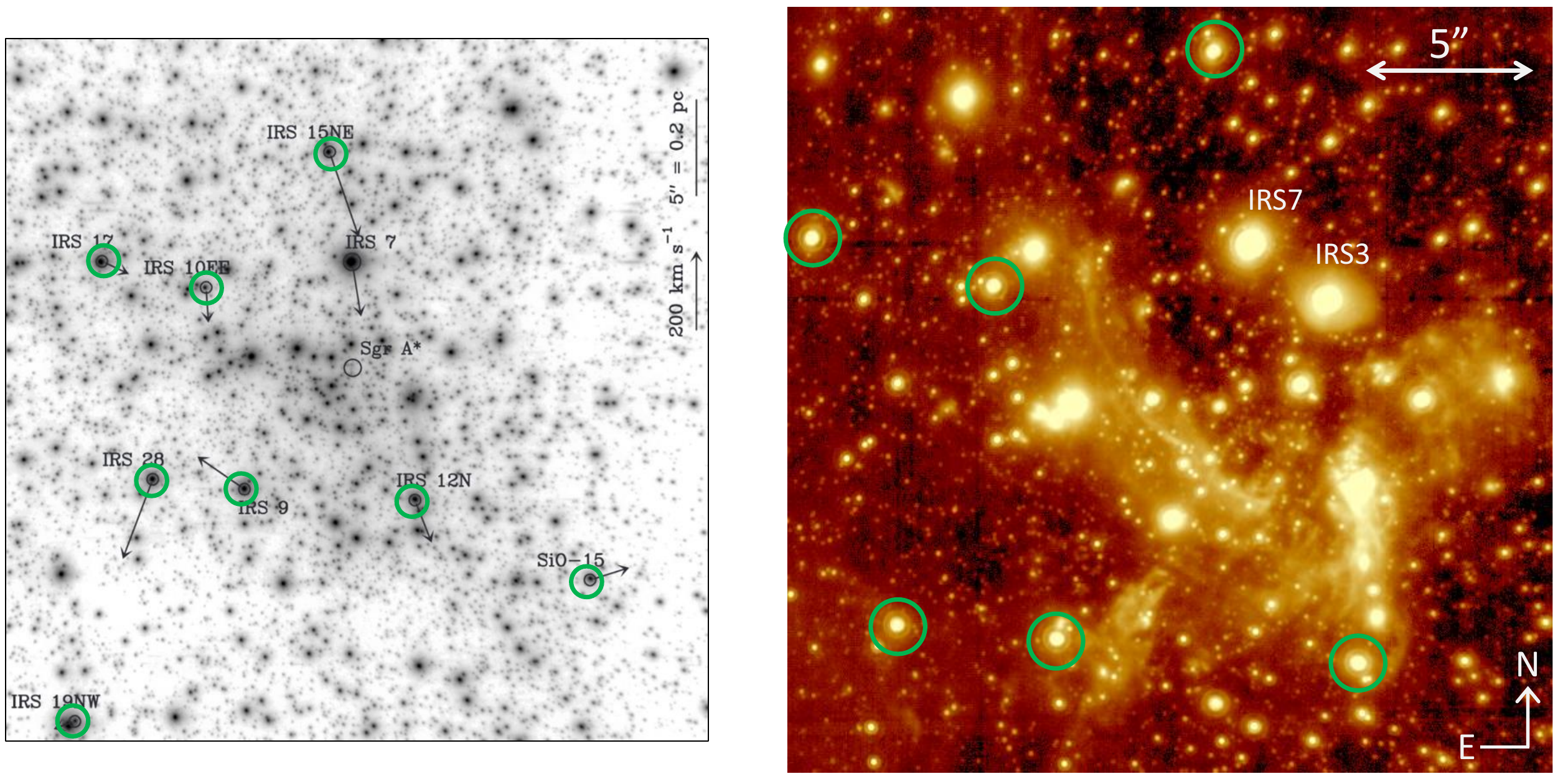}
\caption{L band imaging with NIX of the Galactic Center.
Left: adapted from \citet{rei07} showing the maser stars within 20\arcsec\ of Sgr~A* on a K band image, with their velocities in the plane of the sky. Right: The central 22.8\arcsec\ of the L band NIX image obtained using an autojitter sequence, highlighting stars and warm dust. Diffraction rings can be seen around the brighter more isolated stars, even towards the edge of the field. Six stars for which the corresponding masers are indicated, allow the orientation and plate scale of the camera to be verified.}
\label{fig:gcnix}
\end{figure*}

\paragraph{Galactic Center}

Imaging is an important component to studies of the stellar proper motions in the Galactic Center. Due to the high extinction of $A_V \sim 30$~mag, this can only be done at wavelengths corresponding to H band or longer. Figure~\ref{fig:gcnix} is an image taken through the $L^\prime$ filter which highlights the warm dust features as well as the stars. In this respect, the most striking difference to the K band image is the star IRS3, a very red object cocooned in a dusty shell. The data were taken in the fast uncorrelated read mode of NIX, using the autojitter template to create a random set of 50 offsets within a 15\arcsec\ box.
The exposure time was set to 0.2~sec to avoid saturation of the high thermal background; and at each location 140 individual exposures were taken and averaged together to create a single frame with an equivalent 28~sec integration. This reduces the overhead associated with saving the files while leaving sufficient exposures to efficiently remove the background and compensate the bad pixels.

The same stars and dust feature details are visible as in previous NACO images taken with a similar configuration. The major difference is in the technique used for wavefront sensing: the ERIS data were taken with LGS-AO using an off-axis guide star while the NACO data were taken with an infrared WFS that used IRS\,7, a K=6.5~~mag star less than 6\arcsec\ from Sgr~A*. In this case both datasets were of high quality, and there is little difference between them.

This field has been targeted at radio wavelengths to measure the positions, proper motions, and line of sight velocities of SiO masers \citep{rei07}.
These are shown in the left panel of the figure, and used as an astrometric reference for the infrared imaging.
In this case, it also enables the plate scale and orientation of the \code{LM-13mas} camera to be verified, as given in Table~\ref{tab:cameras}.

\begin{figure}
\centering
\includegraphics[width=\hsize]{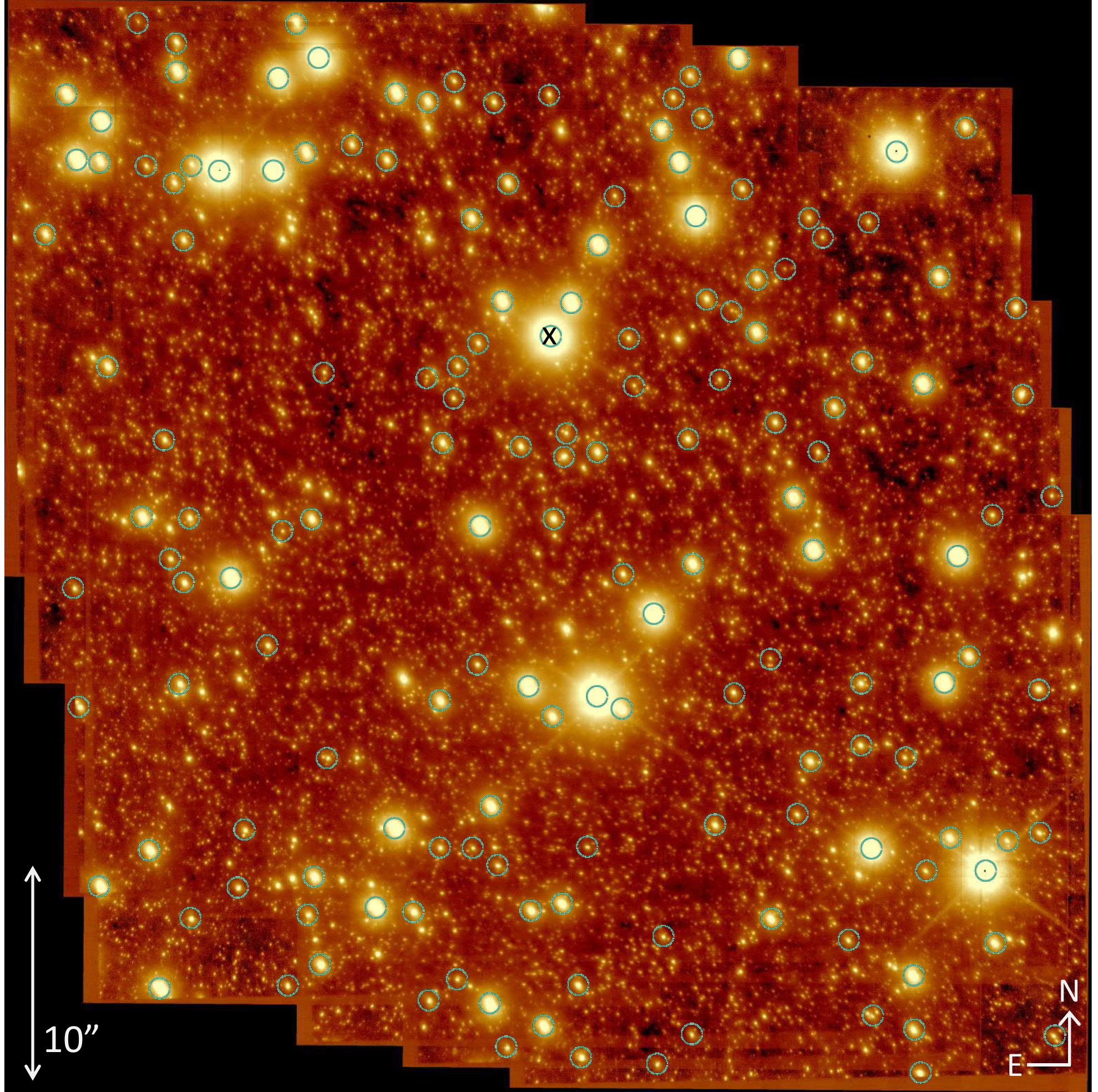}
\caption{K band imaging with NIX of the globular cluster $\omega$~Cen, using LGS-AO for which the star marked by a cross provides tip-tilt and truth sensing. 158 Gaia sufficiently bright and isolated stars are marked in this $5\times5$ mosaic and were used to register the frames as well as derive pixel scale and orientation.}
\label{fig:omegacen}
\end{figure}

\begin{figure*}
\centering
\includegraphics[width=17cm]{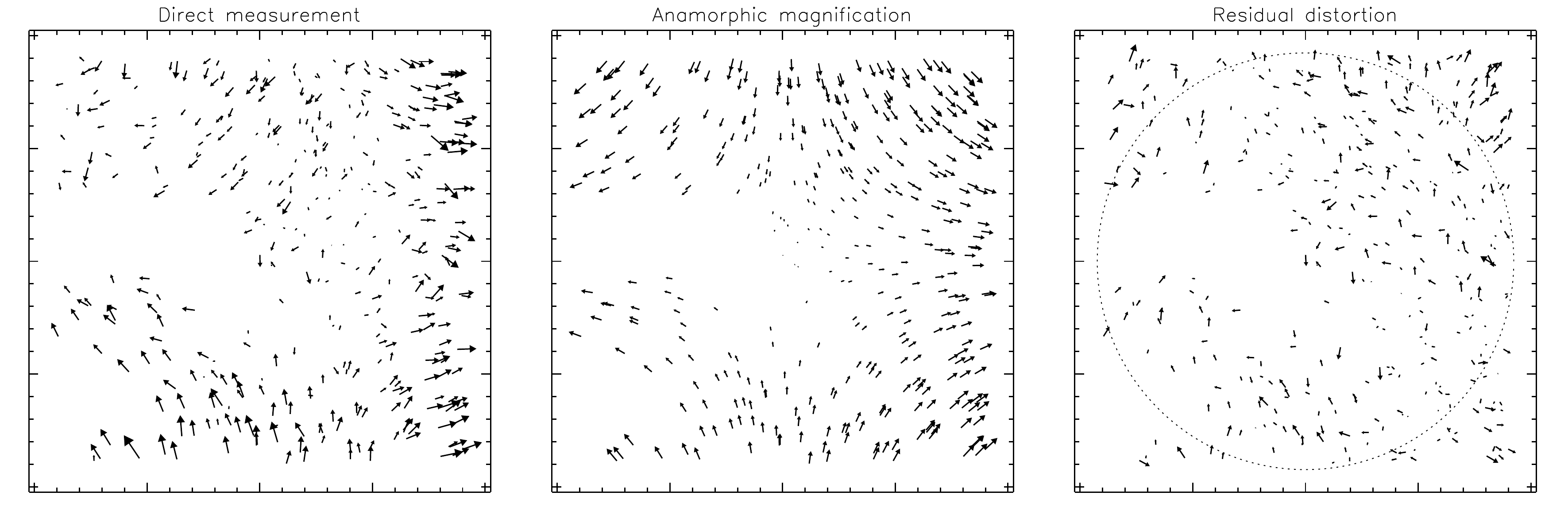}
\caption{Preliminary distortion map of the \code{JHK-27mas} camera derived from the $\omega$~Cen data (that for the \code{JHK-13mas} camera is similar). The dominant term is not from the camera, but comes from the anamorphic magnification due to the high angular tilt of the IR/VIS dichroic. Here, only this linear term has been corrected. Higher order terms are a factor 10 smaller, and below the noise limit of these data. The dotted circle in the right panel represents a 50\arcsec\ aperture where the {\sc rms} residual is only 12~mas even in these noisy data. Note that the centre left region of the map is empty because of the bad pixel cluster.}
\label{fig:distort}
\end{figure*}

\begin{table}[h]
\caption{Basic Geometry of Data for NIX Cameras}
\label{tab:cameras}
\centering
\begin{tabular}{lllll}
\hline\hline
camera & pix scale   & offset    & east \\
       & mas/pix$^a$ & angle$^b$ & direction \\
\hline
\code{JHK-13mas} & 13.09 & $+0.4\degr$ & right \\
\code{JHK-27mas} & 26.92 & $-1.5\degr$ & left \\
\code{LM-13mas}  & 13.03 & $+0.4\degr$ & right \\
\hline
\end{tabular}
\tablefoot{$^a$ pixel scale averaged over all directions. The anamorphism reduces the pixel scale by about 0.2\% in the vertical direction and increases it by the same amount horizontally. $^b$ The offset angle $PA_{offset}$ is defined as the angle (measured east of 'up') that corresponds to north in the detector raw data, for a nominal zero position angle.} 
\end{table}

\paragraph{$\omega$ Cen}

The equivalent verification of plate scale and orientation for the \code{JHK-13mas} and \code{JHK-27mas} cameras was done using the globular cluster $\omega$~Cen, but with a different operational strategy to produce the pipeline reduced image in Fig.~\ref{fig:omegacen}.
The bright star just above centre, marked by a black cross, with $G_{RP} = 11.0$~mag was used as the tip-tilt star in these LGS-AO data -- which have a typical K band FWHM of 100~mas rather than the diffraction limited 55~mas because the LGS-AO performance had not been optimised at this stage of the commissioning.
With the \code{JHK-13mas} camera, a regular grid of $5\times5$ pointings was used, each offset by 5\arcsec\ in one axis and 1\arcsec\ in the other axis. Thus the first and second frames are offset by 5\arcsec\ in right ascension and 1\arcsec\ in declination, while the first and sixth frames are offset by 1\arcsec\ in right ascension and 5\arcsec\ in declination. This allows one to mosaic a larger area, but also ensures that the same regions of the cluster are imaged in very different areas of the detector, enabling an on-sky map of instrumental distortions to be derived.
In addition, applying such large offsets to the object frames with a crowded field provides a stringent test for the precision of the offsets as well as for the pipeline in terms of aligning and combining the data.
A parallel reduction was performed making use of more than 150 stars in the field from the Gaia catalogue \citep{gai22}. 
Similar observations were performed with the \code{JHK-27mas} camera. Both datasets show that the offsets between exposures are accurate to better than 1\% and have a rotational offset of $<$0.1$\degr$.
In addition, they yield the pixel scales and orientations of the two cameras reported in Table~\ref{tab:cameras}.

The instrumental distortions are known to be very small: the NIX camera was specifically designed to have extremely low distortion in the central region, corresponding to $<10$~mas over any $25\arcsec\times25\arcsec$ region, and $<1.5$~mas in the central $7\arcsec\times7\arcsec$ region (in both cases, after a linear coordinate transform has been applied). This is borne out by the ability to register the Galactic Center and $\omega$~Cen frames without needing to apply a distortion correction. A preliminary analysis of the star positions with reference to the Gaia catalogue for this cluster has produced the distortion map in Fig.~\ref{fig:distort}. It indicates that the dominant term is not from the camera, but a correction for the anamorphic magnification due to the high tilt of the IR/VIS dichroic (see Fig.~\ref{fig:censtruct}).
This stretches the data along the columns of the detector by 0.4\%; although because our definition of pixel scale is averaged over all directions, the anamorphism is measured as a vertical stretch combined with a squeeze of the data horizontally, both by 0.2\% with respect to a nominal square field.
After correcting for this linear term, the residuals in the data analysed here are dominated by noise. 
Nevertheless the {\sc rms} distortion in the central 25\arcsec\ of the \code{JHK-13mas} camera is reduced to 10~mas. Similarly, as illustrated in  Fig.~\ref{fig:distort}, within the central 50\arcsec\ aperture of the \code{JHK-27mas} camera the residual {\sc rms} distortion is measured to be 12~mas.
Optical modelling of the instrument indicates that higher order distortion terms are more than a factor of 10 smaller than the anamorphic magnification.

\begin{figure*}
\centering
\includegraphics[width=17cm]{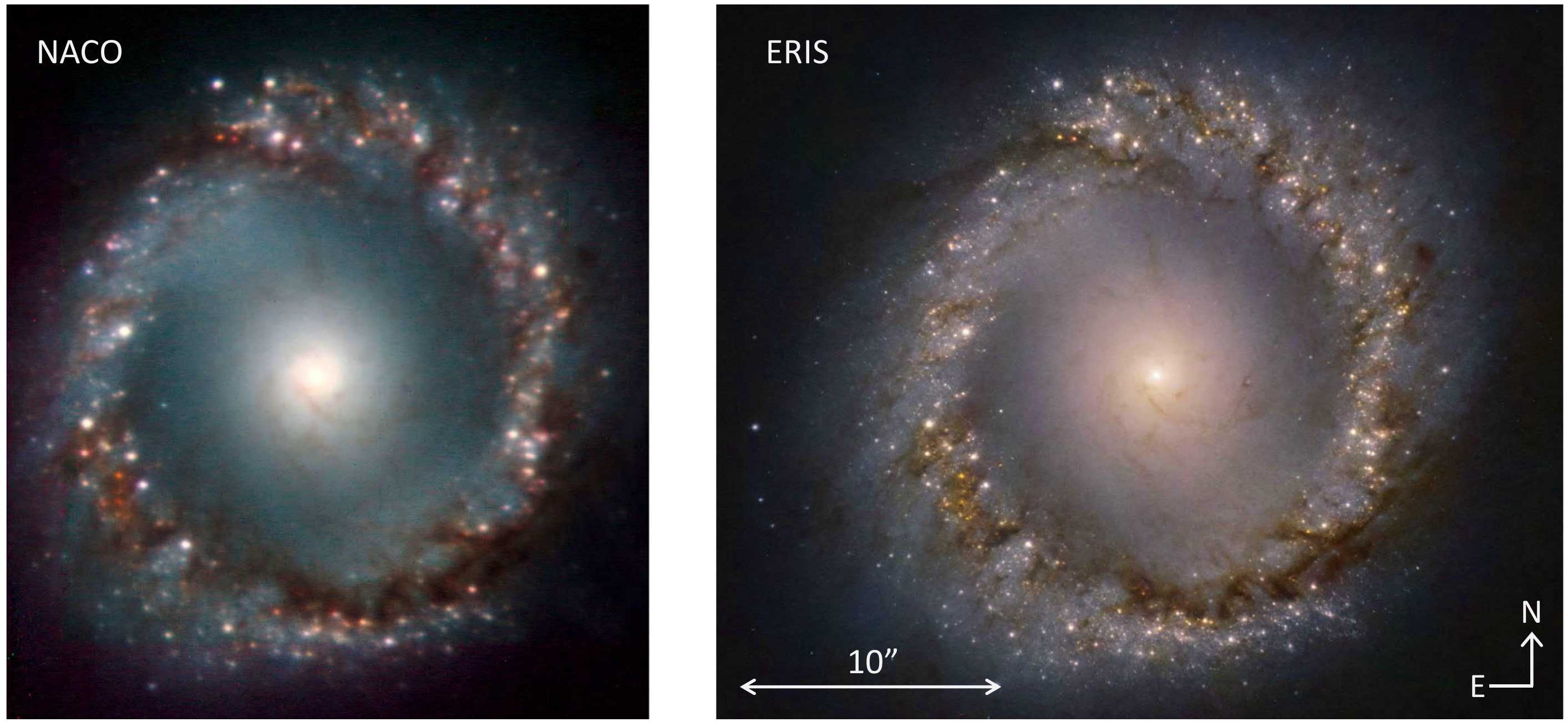}
\caption{Pseudo-colour images of the circumnuclear ring in NGC~1097, created from J, H, and K images represented by blue, green, and red. Comparison of the NACO image (Fig.~1 of \citealt{pri05}; credit: ESO/Prieto et al.) to the ERIS image (credit: ESO/ERIS team) reflects the improvement in adaptive optics technology over the last 20 years.}
\label{fig:ngc1097}
\end{figure*}

\paragraph{NGC1097}

This active galaxy has previously been targeted with both NACO \citep{pri05} and SINFONI \citep{dav09}. It has a prominent star forming circumnuclear ring within 1~kpc of the nucleus. On large scales, a strong bar is channelling gas in to the ring; and on smaller scales a circumnuclear spiral is driving gas further in to the nucleus.
It is challenging for adaptive optics because the nucleus, although point-like and moderately bright in the optical, is superimposed on a bright background corresponding to the centre of the galaxy.
The NACO observations were obtained in 2002, closing the high order adaptive optics loop on the nucleus itself.
\citet{pri05} reported resolutions of $0\farcs18\times0\farcs15$ in K band to $0\farcs20\times0\farcs19$ in J band. These were measured from the nucleus as well as individual H{\sc ii} regions in the ring, which were resolved by HST to have sizes of $\sim4$~pc ($0\farcs06$) and thus do not affect the resolution measurement in NACO.
Anisoplanatism over the 10\arcsec\ distance from the nucleus to the ring is also not expected to have a strong impact on the resolution measurement (although it would reduce the Strehl ratio).
With ERIS, LGS-AO was used, with the nucleus acting as the tip-tilt reference. After removing the smooth background contribution, the resolution was measured from the nucleus to be 77~mas and 85~mas in J and K band respectively: a factor 2-2.5 better than NACO.
In the ring, the 50 brightest H{\sc ii} regions were measured to be symmetric and have sizes of 92~mas in J band and 98~mas in K band.
Correcting for their intrinsic size leads to resolution estimates of $\sim70$~mas and $\sim78$~mas respectively, which are comparable to those derived from the nucleus.
This illustrates a different regime from those described earlier, where ERIS can exploit the robust laser guide star system of the AOF to achieve significantly better performance than has been possible previously with only NGS-AO.

\subsection{High Contrast Imaging}

\begin{figure*}
\centering
\includegraphics[width=16cm]{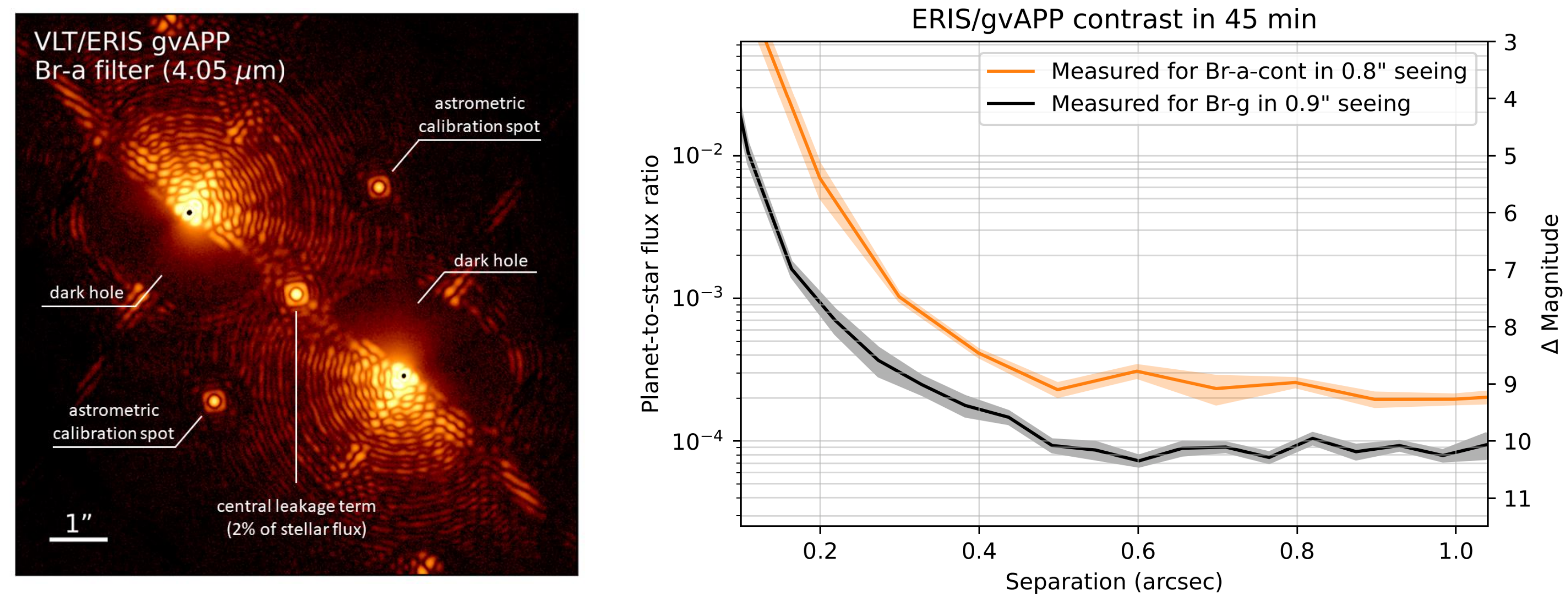}
\caption{On-sky performance of the gvAPP in NIX. Left: image of the PSF obtained with the mask through the \code{Br-a} filter, which matches the theoretical PSF shown in Fig.~\ref{fig:nixapp}. The image is drawn on a logarithmic scale, and the key elements are labelled.
Right: Contrast curves for the narrow \code{Br-a-cont} and \code{Br-g} filters in the L band and K band respectively. The \code{Br-a-cont} contrast curve is retrieved using a fake planet injection and recovery test on 45~min of ADI/PCA processed data of HR~8799. The same procedure is applied to a 45~min \code{Br-g} dataset of $\gamma$~Gru. The shaded areas trace the median absolute deviation of the contrast dependent on the parallactic angle of the injected fake planet. We note that the achievable contrast depends on many factors including star magnitude, filter bandwidth and wavelength, seeing, integration time, and data processing method.}
\label{fig:appcontrast}
\end{figure*}

Among the high contrast modes, the gvAPP is relatively straightforward to use because the mask is located in the pupil plane and so, unlike the FPC, there is no need to provide feedback from the science image to keep the mask and star co-aligned.  The operation of the FPC is somewhat more complex because, to reach optimal performance, it requires an active fine tuning of the pointing of the star onto the vortex mask center to correct for slow drift (due to e.g. mechanical flexure, or small motions occurring in pupil tracking mode). While contrast curves for the FPC are still to be finalized, good pointing stability after QACITS closed-loop control has been achieved with centering errors below 0.025~$\lambda/D$ at a correction rate of 0.2~Hz. The preliminary on-sky performance of the FPC show broadband raw null depths better than 1:100. 

Data from the gvAPP and FPC modes also provide essential tests for the pupil tracking mode, including the ability to offset to repeatable locations on the detector while the sky coordinates are rotating, and for the AO system to compensate the differential motion between the guide star position and the science camera pointing. The on-sky image from the gvAPP in the left panel of Fig.~\ref{fig:appcontrast} matches the theoretical prediction in Fig.~\ref{fig:nixapp} extremely well in terms of the two half-PSFs with dark holes, central leakage term for photometry, and the two astrometric calibration spots.
This image is of the star HD~203387 ($\iota$~Cap), a variable G6 to G8 giant with a mass of 3~$M_\odot$ at a distance of 62~pc \citep{cho95,hen95}.
It is taken through the \code{Br-a} filter at 4.05~$\mu$m using the \code{LM-13mas} camera.
Two sets of data were taken on this object, in both cases using 3~sec exposures with the slow read mode.
The first set were taken in $0\farcs6$ seeing, and comprises 12 groups of 25 exposures that were taken in two different locations on the detector separated by $7\farcs9$. During these observations there was relatively little field rotation.
The second set, shown here, comprises 11 groups of 50 exposures, taken in seeing that varied from $0\farcs7$ to $1\farcs1$.
They were also dithered between two locations on the detector, and the background was derived simply by subtraction of the opposite positions.
The offset used was comparable but since there was 20$\degr$ field rotation during the observations, the actual offsets on-sky varied from $3\farcs2 - 4\farcs1$ in right ascension and $7\farcs1 - 7\farcs5$ in declination.

Recently, based on analysis of radial velocity and astrometric data, two exoplanets have been reported around HD~203387 \citep{fen22}.
These have masses of 3.5~$M_J$ and 7.5~$M_J$ and semi-major axes of 30~mas and 80~mas respectively, and as such are too close to the primary to image directly in L band. 
However, a candidate companion was detected at a projected distance of 1\farcs6, which is outside the dark zone.
The contrast is 8.3~mag with respect to the $L^\prime = 2.2$~mag primary star, suggesting it is a low mass M dwarf.
The field rotation during the observations confirms the detection, but without measurements at other epochs we cannot rule out that this is simply a background star.

The indicative contrast one might expect to reach with the gvAPP is shown in the right panel of Fig.~\ref{fig:appcontrast} for two particular filters -- noting that the achievable contrast depends on many things including the AO performance (star magnitude and seeing), the noise (filter bandwidth and wavelength, integration time, field rotation) and the data processing method applied. To calculate the contrast curves, the data are first pre-processed using Pynpoint \citep{sto19,ama12}, where the two dark zones of the gvAPP PSF are stitched together to form a single science cube. Then, a fake planet injection and retrieval is performed using the HCI statistics software Applefy \citep{bon23} and utilising PCA/ADI for the subtraction of the stellar PSF. The unsaturated gvAPP PSF is used as a PSF reference for the fake planet. The number of PCA components is varied and chosen independently for different separations of the injected fake planet. 
%

\subsection{Sparse Aperture Masking}

\begin{figure}
\centering
\includegraphics[width=\hsize]{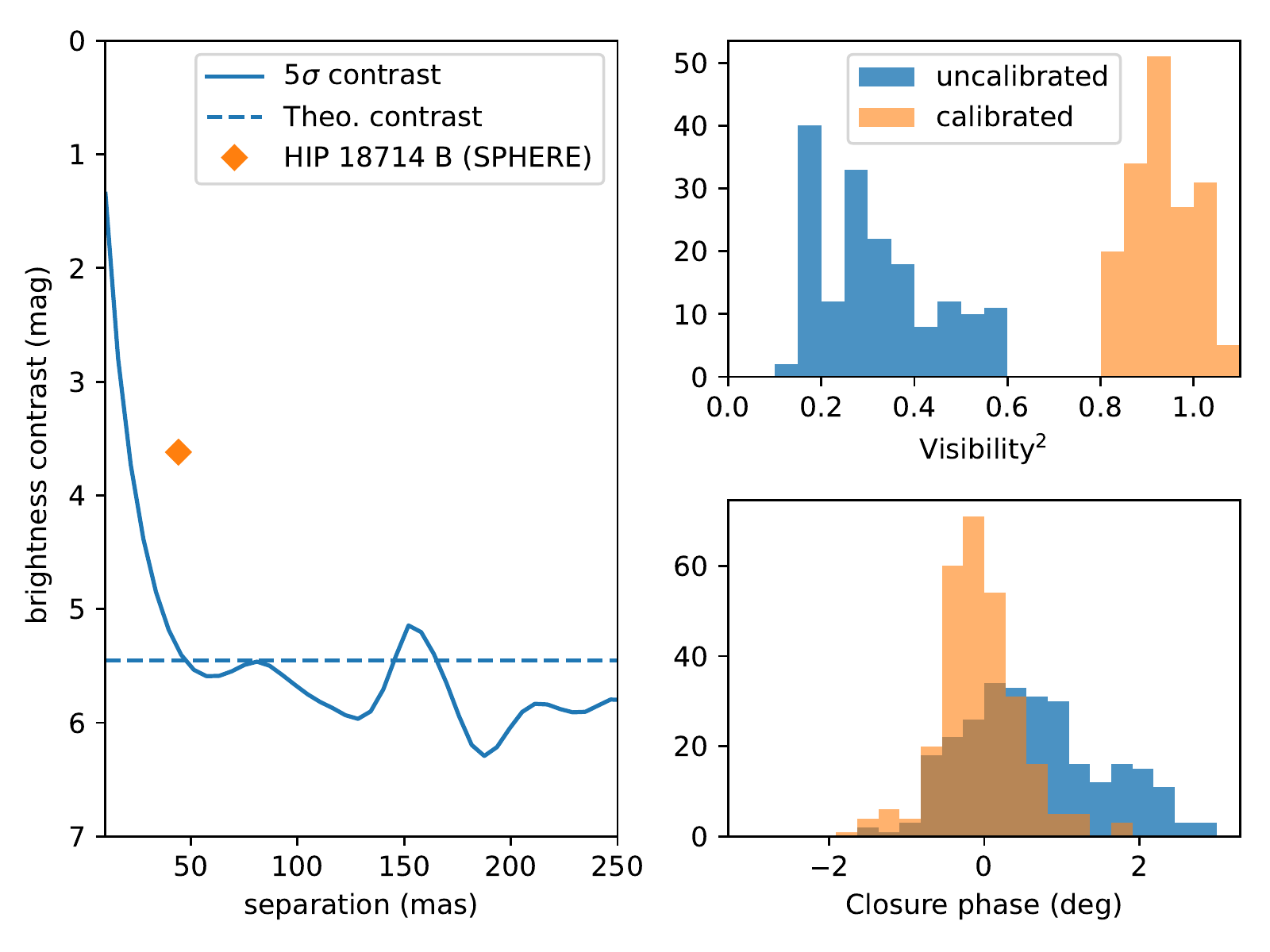}
\caption{On-sky performance of the SAM-7 mask. Left: the solid line is the $5\sigma$ contrast curve observed on HIP\,18714 with the mask and the \code{Ks} filter. The diamond marker shows companion as detected by SPHERE in the J band \citep{2022A&A...663A.144B}, but not detected by us. The dashed line corresponds to the $5\,\sigma$ theoretical dynamic range assuming an accuracy of 0.5 degrees for the closure phase \citep[according to Eq.\ (2) in][]{2011A&A...532A..72L}.
Right: Histograms of the visibility squared and closure phase per exposure. The standard deviation of the visibity and phase after calibration are respectively 0.06 and 0.52 degrees.
}
\label{fig:sam_hip18714}
\end{figure}

The aperture masking mode of ERIS is a convenient tool for observing systems at very small angular separations. Thanks to its interferometric nature, it is possible to adjust models of the targets to obtain spatial information below the limit of diffraction. HIP\,18714 (HD\,25402A) was observed in 2017 by \citet{2022A&A...663A.144B} using the SPHERE instrument in the J band. They found a very close companion at a separation of 44 mas (2.3 au) and a position angle of 351.6 degrees. They estimated the mass to be 0.25 solar masses (0.23 in mass ratio). Additionally, HIP\,18714 is part of the \texttt{nss\_acceleration\_astro} Gaia DR3 archive table \citep{2022arXiv220605726H}. The proper motion is $69.09\pm0.07$\,mas in RA, $-5.21\pm0.06$\,mas in DEC. It is a member of the Tucana-Horologium association according to BANYAN $\Sigma$ classification \citep{2018ApJ...856...23G}: >80\% including the RV, and >99\% without. The acceleration of the star ($-1.44\pm0.08$\,mas/yr$^2$ in RA, $-0.54\pm0.08$\,mas/yr$^2$ in Dec) as well as the derivative of the acceleration ($1.4\pm0.4$\,mas/yr$^3$ in RA, $7.0\pm0.3$\,mas/yr$^3$ in Dec) also point toward the presence of a close companion.

During the night of November 5$^{\rm th}$, 2022, exactly 5 years after the SPHERE observations, we observed HIP\,18714 on ERIS with NIX and the SAM 7-hole pupil mask and the Ks filter. We spent a total time on target of 30 minutes, followed by 15 minutes on the calibrator HD\,26072, for a total duration of 1 hour (including overheads and acquisition). The science observation consisted of 8 exposures of 200 frames of 0.8 seconds. Between two exposures, the diffraction pattern was nodded to one of 4 different positions on the detector. The calibrator was also carefully nodded to the same positions, and observations consisted of 4 exposures with the same integration times. The diffraction patterns were background subtracted using the images at other nodding positions, cropped, and bad pixels corrected. The data were reduced using the AMICAL software \citep{2020SPIE11446E..11S} to produce oifits files. The HIP\,18714 oifits were calibrated using the HD\,26072 oifits 
\citep[angular diameter of 0.3\,mas according to][]{2019MNRAS.490.3158C}.

There is a direct correlation between the dynamic range and the accuracy of the closure phase. In our dataset, the statistical deviations observed between frames within one exposure are approximately 1.0 degree. However, after calibration and averaging over the frames, the deviation to zero is still $\sigma(CP)=0.52$ degrees per exposure, which is higher than the expected deviation of less than 0.1 degree if the target was unresolved and no bias existed. According to Eq.\ (2) from \citet{2011A&A...532A..72L}, this accuracy gives an upper limit to the flux ratio of a companion to $\sigma(CP) \times 2.5\times 10^{-3}=1.3 \times 10^{-3} $ at 1 sigma. The visibility squares were also reduced, and after calibration, they averaged at $0.95\pm0.6$.

We fitted a model of a binary star to the closure phase. The fit output is a $\chi^2$ map as a function of ascension and declination of the secondary. It shows many local minima, which do not suggest a significant binary solution. The best solution indicates a flux ratio of $3\times 10^{-3}$ (6.3 magnitudes) at 43 mas. This solution is much fainter than the SPHERE prediction, which was 3.62 magnitudes in the J band. This suggests a non-detection of the companion. Indeed, it can be demonstrated that the companion would have been detected at 5$\sigma$ at any separation above 20\,mas (as shown in the left panel of Fig.~\ref{fig:sam_hip18714}). This means that the stellar companion would have moved South by 24 mas within 5 years. With a mass ratio of 0.23 (as per Bonavita et al.), this would result in a proper motion induced to the star of $\approx 1$\,mas/yr. With a derivative to the acceleration of 7\,mas/yr$^3$ in declination, this is entirely possible. A more rigorous orbital fitting would nevertheless be necessary to confirm this hypothesis. Last, it is worth noting that the companion was not detected in 2016, on the Gemini Multi-Object Spectrograph (GMOS) by \citet{2018AJ....156..137B}.

\subsection{Longslit Spectroscopy}

\begin{figure*}
\centering
\includegraphics[width=15cm]{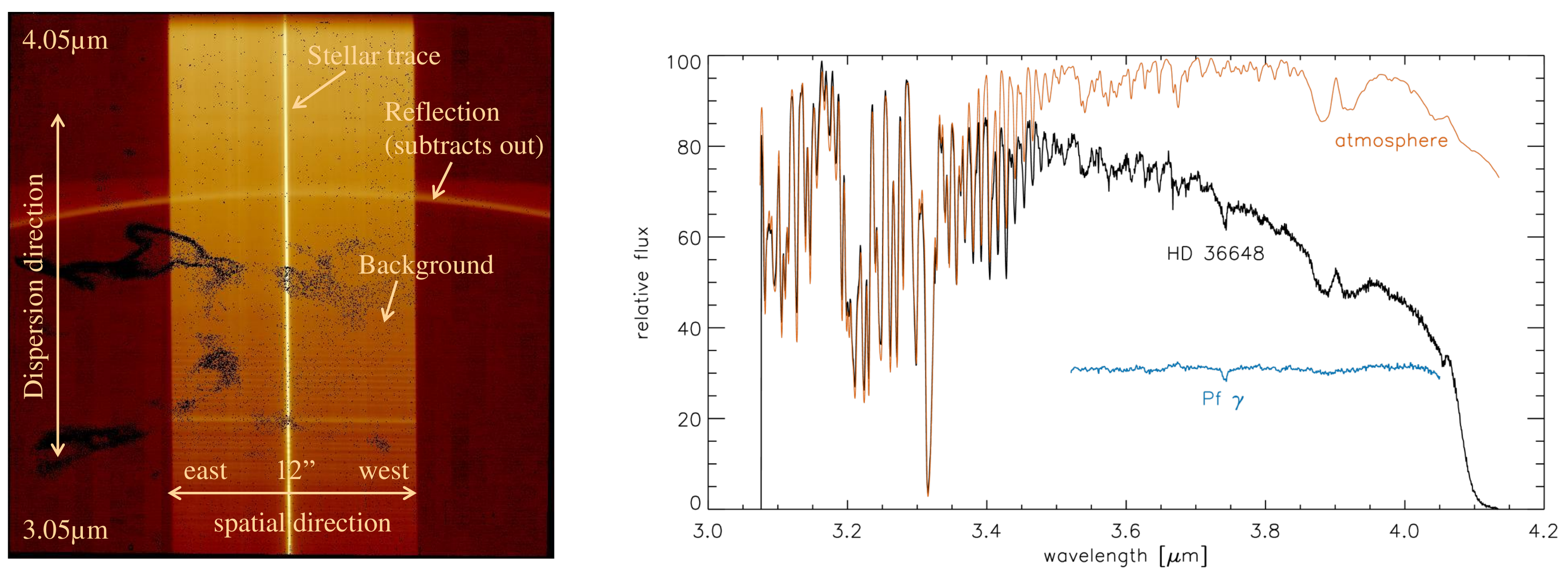}
\caption{On-sky data obtained with the long slit spectrograph in NIX. Left: image of a raw 2D L band spectrum, with the various features labelled (the east side of the slit is indicated for an on-sky position angle of 0\degr).
Right: an extracted spectrum of the double star HD~36648 together with an atmospheric transmission profile taken from the Molecfit library. The rapidly varying transmission at $<3.5~\mu$m makes it hard to correct the spectrum that region. The blue spectrum in the cleaner part, which has had the atmospheric features removed and its slope normalised, shows the detection of Pf\,$\gamma$ at 3.74~$\mu$m, as expected given the A5 spectral type of this star.}
\label{fig:lss_hd36648}
\end{figure*}

The application of slit spectroscopy in ERIS is no different to the typical operation in many other instruments. 
Centering of the PSF in the slit, although a standard procedure, needs to be done carefully because its FWHM is comparable to the narrow 86~mas slit width.
The steps include imaging the source and subtracting as background an image of the slit; but it is aided by the fact that the slit is aligned along the rows of the detector, and the offsetting directions are defined parallel and perpendicular to the slit independent of the position angle on-sky to which it is set.

The appearance of the raw data is illustrated in the left panel of Fig.~\ref{fig:lss_hd36648}.
The 12\arcsec\ slit is oriented horizontally, and the dispersion direction is close to vertical.
There is some scattered light, but this is constant and subtracts out well.
The central part of the spectrum can be affected by bad pixels, particularly for locations to the left side of the slit; and so the nodding positions should be selected accordingly.

An example spectrum, for which the template was configured to nod the star to four positions along the slit and take a 20~sec exposure each time, is shown in the right panel of Fig.~\ref{fig:lss_hd36648}.
The spectrum is of HD~46638, a 6.5~mag double star with 7\arcsec\ separation, and having A5 spectral type \citep{can93}.
Superimposed is the atmospheric transmission profile taken from the spectral library associated with Molecfit \citep{sme15}.
This shows that at $\lambda \la 3.5~\mu$m, the transmission varies strongly and rapidly with wavelength -- making it hard to correct, in particular because the spectral profile has imprinted on it the illumination of the source across the width of the slit.
This varies by a factor two from the edge to the centre of the 86~mas slit width because the spatial profile of the diffraction limited PSF has a FWHM of about 90~mas, which is comparable to the slit width.
In the cleaner half of the bandpass, the final spectrum, for which the slope was normalised, shows that correction with an atmospheric template works well.
It reveals an absorption feature with equivalent width of 0.33~\AA\ corresponding to Pf\,$\gamma$ at 3.741~$\mu$m.
The line width is measured to be 730~km~s$^{-1}$.
If we assume it is unresolved, it indicates a spectral resolution of $R = 970$ consistent with that expected as indicated in Sec.~\ref{sec:nix}.


\section{Conclusions}
\label{sec:conc}

ERIS is a new instrument for the VLT, with adaptive optics capability that makes use of the Adaptive Optics Facility on UT4, and allows one to perform diffraction limited observations using guide stars currently as faint as $G_{RP} = 18$~mag. It replaces, and enhances the capabilities of, two previous instruments, offering a variety of observing modes:
\begin{itemize}
\item integral field spectroscopy at $1-2.5~\mu$m at resolutions of $R\sim5000$ and $R\sim10000$ (not matched by any other ground or space-based near-infrared IFU) over a field of view from 0\farcs8 to 8\farcs0;
\item imaging at $1-5~\mu$m over fields of almost 30\arcsec and 60\arcsec;
\item high contrast imaging using focal and pupil plane masks, and sparse aperture masks;
\item long slit spectroscopy at $3-4~\mu$m.
\end{itemize}
The instrument was first offered for open time in ESO's Observing Period 111, which started in April 2023.

For those planning to observe with ERIS, it is essential to read the User Manual\footnote{\href{https://www.eso.org/sci/facilities/paranal/instruments/eris.html}{https://www.eso.org/sci/facilities/paranal/instruments/eris.html}}, 
which provides up-to-date information about the instrument and its operational modes, as well as more specific information relevant to the preparation of observations.
Similarly, sensitivities of the various modes in differing atmospheric conditions and AO performance are best assessed using the online Exposure Time Calculator\footnote{\href{https://www.eso.org/observing/etc}{https://www.eso.org/observing/etc}}. 
ERIS/SPIFFIER and ERIS/NIX have advanced data processing pipelines developed by the ERIS consortia and ESO.
Both pipelines make use of the ESO Reflex user interface environment \citep{fre13} and are available for download\footnote{\href{https://www.eso.org/sci/software/pipelines}{https://www.eso.org/sci/software/pipelines}}.

\begin{acknowledgements}

The consortium would like to thank the many people who have contributed to the project in numerous different ways, in some cases `behind the scenes', both within the partner institutes as well as at ESO in Garching and at Paranal.
The authors would also like to acknowledge the help of J. Kammerer, E. Huby, T. Stolker, and J. Sanchez for preparing and reducing the SAM observations of HIP\,18714.
JH gratefully acknowledges the financial support from the Swiss National Science Foundation under project grant number $200020\_200399$.
The sparse aperture masking mode has received funding from the French National Research Agency (ANR-13-JS05-0005) and from the European Research Council (ERC) under the Horizon 2020 research and innovation programme (Grant agreement No. 639248).
The research of FS leading to these results has received funding from the European Research Council under ERC Starting Grant agreement 678194 (FALCONER)
Parts of this work have been carried out within the framework of the National Centre of Competence in Research PlanetS supported by the Swiss National Science Foundation under grants 51NF40\_182901 and 51NF40\_205606. S.P.Q. and A.M.G. acknowledge the financial support of the SNSF.
The development of the FPC mode has received funding from the European Research Council (ERC) under the European Union’s Horizon 2020 research and innovation programme (grant agreement No 819155), and from the Wallonia-Brussels Federation (grant for Concerted Research Actions).
The paper makes use of public data released from the ERIS commissioning observations at the VLT Yepun (UT4) telescope under Programmes ID 60.A-9917.

\end{acknowledgements}

%
%

\bibliographystyle{aa}
\bibliography{eris2022}

\end{document}